\documentclass[twocolumn,amsmath,amssymb,aps,prb]{revtex4-1}

\pdfoutput=1

\usepackage{graphicx}

\usepackage{sidecap}
\usepackage{multirow}
\usepackage[mathcal]{euscript}
\usepackage{epsf, times}
\usepackage{subfigure}

\usepackage{amssymb,amsfonts,amsmath}
\usepackage{dcolumn} 
\usepackage{bm, bbm}      
\usepackage{curves}
\usepackage{epic}
\usepackage{wasysym}
\usepackage{epsf, times}
\usepackage{subfigure}
\usepackage{amsthm}

\newcommand{\beq}{\begin{equation}}
\newcommand{\eeq}{\end{equation}}
\newcommand{\bea}{\begin{eqnarray}}
\newcommand{\eea}{\end{eqnarray}}
\def\DTO{Dy$_2$Ti$_2$O$_7$}

\begin{document}

\title{Tunable non-equilibrium dynamics: field quenches in spin ice}

\author{S.~Mostame$^{1,3}$, C.~Castelnovo$^2$, R.~Moessner$^3$, and S.~L.~Sondhi$^4$}
\affiliation{
$^1$ Department of Chemistry and Chemical Biology, 
Harvard University, 
Cambridge, MA 02138, USA
      }
\affiliation{
$^2$ TCM group, Cavendish Laboratory, 
University of Cambridge, 
Cambridge, CB3 0HE, United Kingdom
      }
\affiliation{
$^3$ Max Planck Institute for the Physics of Complex Systems, 
N\"{o}thnitzer Str. 38, D-01187 Dresden, Germany
      }
\affiliation{
$^4$ Department of Physics, 
Princeton University, 
Princeton, NJ 08544, USA 
      }

\date{\today}

\begin{abstract}
We present non-equilibrium physics in spin ice as a novel setting which 
combines kinematic constraints, emergent topological defects, and 
{\it magnetic} long range Coulomb interactions. 
In spin ice, magnetic frustration leads to highly degenerate yet 
locally constrained ground states. Together, they form a highly unusual 
magnetic state -- a ``Coulomb phase'' -- whose excitations are pointlike 
defects -- magnetic monopoles -- in the absence of which effectively no 
dynamics is possible. Hence, when they are sparse at low temperature, 
dynamics becomes very sluggish.
When quenching the system from a monopole-rich to a monopole-poor state, a 
wealth of dynamical phenomena occur the exposition of which
is the subject of this article. Most notably, we find reaction diffusion 
behaviour, slow dynamics due to kinematic constraints, as well as a regime 
corresponding to the deposition 
of {\it interacting} dimers on a honeycomb lattice. We also identify new 
potential avenues for detecting the magnetic monopoles in a regime of 
slow-moving monopoles. 
The interest in this model system is further enhanced by its large degree of 
tunability, and the ease of probing it in experiment: 
with varying magnetic fields at different temperatures, geometric properties 
-- including even the effective dimensionality of the system -- can be varied. 
By monitoring magnetisation, spin correlations or zero-field Nuclear Magnetic 
Resonance, the dynamical properties of the system can be extracted in 
considerable detail. This establishes spin ice as a laboratory of choice for 
the study of tunable, slow dynamics.
\end{abstract}
%
%

\maketitle

\section{Introduction}

Nature and origin of unusual -- in particular, slow -- dynamics 
in disorder-free systems~\cite{Stinchcombe2001} are amongst the most 
fascinating aspects of disciplines as diverse as the physics of structural 
glasses and polymers~\cite{Angell1995}, chemical reactions and biological 
matter~\cite{Tokuyama2004}. 
Kinetically constrained models, following the original idea by Fredrickson 
and Andersen~\cite{Fredrickson1984}, represent one paradigm in which unusual 
dynamics is generated by short-distance ingredients alone without 
disorder~\cite{Ritort2003}. 
Another is provided by reaction-diffusion systems, in which spatial and 
temporal fluctuations feed off each other to provide a wide variety of 
dynamical phenomena~\cite{Toussaint1983} especially due to the slow decay of 
long wavelength fluctuations. 

Spin ice systems~\cite{Bramwell2001} allow to combine both aspects -- 
thanks to the nature of their emergent 
topological excitations, which take the form of magnetic 
monopoles~\cite{Castelnovo2008,Castelnovo2012} with long range Coulomb 
interactions. 
The ground state correlations in these localised spin systems lead to 
kinematic constraints in the reaction-diffusion behaviour of these mobile 
excitations~\cite{Castelnovo2010,Levis2012}. 

Understanding the dynamics of spin ice systems, and in particular proposing 
new ways to probe their out-of-equilibrium properties, is of direct 
experimental 
relevance. For instance, modelling the emergent excitations near 
equilibrium~\cite{Ryzhkin2005,Jaubert2009} allowed to gain insight 
on the observed spin freezing at low 
temperatures~\cite{Matsuhira2000,Snyder2004}, 
and to explain in part the time scales measured in 
zero-field NMR~\cite{Henley2013}. 
Despite the fast paced progress, several open questions remain unanswered in 
particular concerning the behaviour of spin ice materials far from equilibrium, 
as evidenced for instance by recent low-temperature magnetic relaxation 
experiments~\cite{Matsuhira2011,Yaraskavitch2012,Revell2012}. 

In this article, we study the strongly out-of-equilibrium behaviour 
following a quench from a monopole-rich to a monopole-poor regime by 
means of varying an applied magnetic field. 
We uncover a wide range of dynamical regimes and we provide a theoretical 
understanding of their origin. We show that the initial evolution maps onto 
a deposition of dimers on a honeycomb lattice, in presence of long range 
Coulomb interactions between the `vacancies', i.e., the uncovered sites. 
The long time behaviour can instead be understood as a dynamical arrest due to 
the appearance of field-induced energy barriers to monopole motion. 
This regime can be seen as similar to conventional spin ice, but with a 
monopole hopping time exponentially sensitive to temperature: we have a Coulomb 
liquid in which time can pass arbitrarily slowly. 

We discuss how to probe these phenomena in experiment, showing how by 
monitoring magnetisation, spin correlations or 
zero-field Nuclear Magnetic Resonance (NMR), the dynamical properties of the 
system can be extracted in detail. 

Overall, this richness and versatility establishes spin ice as a 
laboratory of choice for the study of slow dynamics arising from an interplay 
of frustration (local constraints), topological defects (monopoles) and 
magnetic long-range Coulomb interactions.
%
%

\section{
Phase diagram and quench protocols
        }
Spin ice materials consist of magnetic rare earth ions 
arranged on a corner-sharing tetrahedral (pyrochlore) lattice. 
Their magnetic moments are well-modeled by classical Ising spins  
constrained to point in the local $[111]$ direction (either into or out of 
their tetrahedra)~\cite{Bramwell2001}, as illustrated in the top left panel 
of Fig.~\ref{fig: three phases} (see also the Suppl. Info.). 

The frustration of the magnetic interactions manifests itself in the fact that 
there are not just a handful of ground states -- an unfrustrated Ising 
ferromagnet, for instance, has only two ground states. 
Rather, spin ice has an exponentially large number of ground states, namely 
all those configurations which obey the {\em ice rule} that each tetrahedron 
has two spins pointing into it, and two out (2in-2out). 

At any finite temperature, there will be excitations in the form of 
tetrahedra which violate the ice rule, 
having three spins pointing in and one out or vice versa. 
Crucially, these defects can be thought of as {\em magnetic monopoles} which 
are deconfined and free to carry their magnetic charge across the 
system~\cite{Castelnovo2008}. 
The energy cost $\Delta$ associated with the creation of a monopole is 
of the order of a few degrees Kelvin for spin ice materials Dy$_2$Ti$_2$O$_7$ 
and Ho$_2$Ti$_2$O$_7$, leading to an exponential suppression 
of their density for $T \ll \Delta$. 

We refer the reader interested in the detailed background to 
Ref.~\cite{Castelnovo2012,Henley2010}. 
Here, we emphasize that spin ice is the only experimentally accessible 
magnetic system that realises a Coulomb phase described 
by a low-energy emergent gauge field. The new phenomena we describe here can 
ultimately be traced back to this fundamentally new feature.
%
%

\subsection{
The equilibrium phase diagram of spin ice
           }
\begin{figure}
\includegraphics[]
                {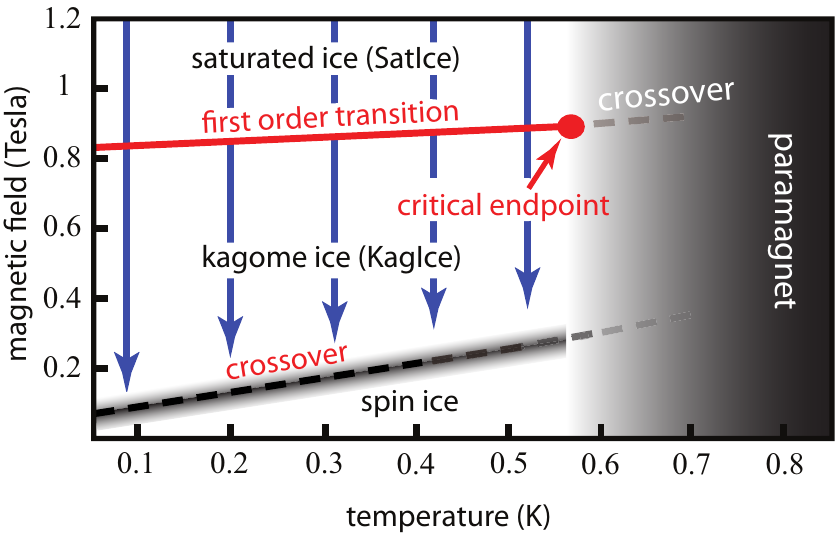}
\caption{\label{fig: 111 phase diagram} 
Sketch of the equilibrium phases of spin ice in presence 
of a $[111]$ magnetic field. 
The blue vertical arrows denote examples of field quenches discussed in 
this work.
}
\end{figure}

\begin{figure*}
%
\hspace{0.2 cm}%
\includegraphics[]
                {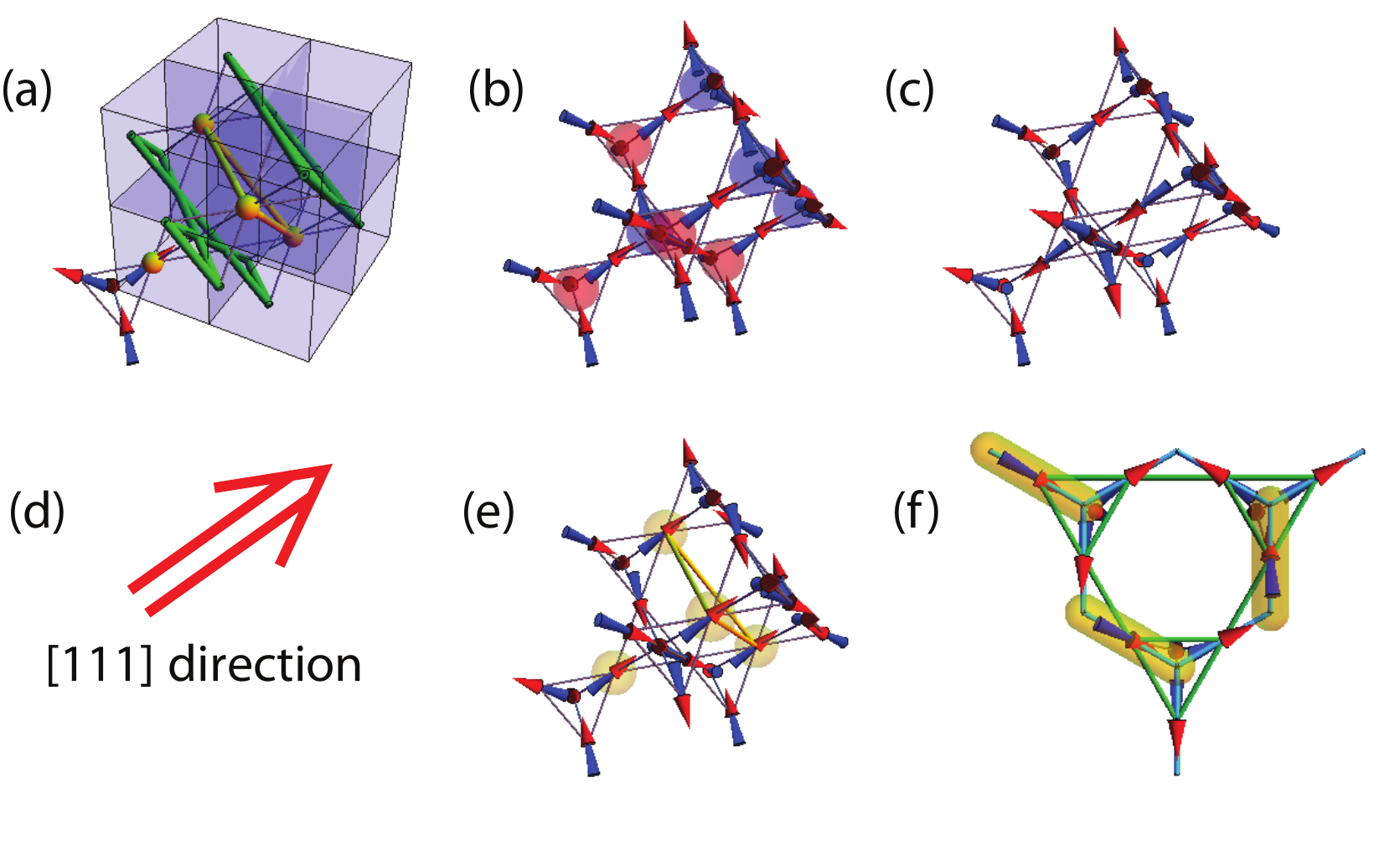}
\caption{\label{fig: three phases} 
(a) 
Pyrochlore lattice formed by the rare earth ions, 
highlighting the layered structure of alternating triangular and kagome 
planes in the $[111]$ crystallographic direction; 
the four spins in the bottom left tetrahedron illustrate the local 
easy axes and the 2in-2out ice rule. 
(b) 
Saturated ice state (SatIce), where all the spins have a positive 
projection along the $[111]$ field direction (d); each tetrahedron hosts a 
monopole (positive 3in-1out shaded in red, and negative 3out-1in shaded 
in blue) forming a crystal of ``magnetic ions''. 
If we consider for instance the two topmost tetrahedra, they host a 
positive (3in-1out) and a negative (1in-3out) monopole, respectively; 
the pair is \textit{non-contractible}, since reversing the intervening spin 
does not annihilate the monopoles but rather produces an even higher energy 
excitation (a pair of 4in and 4out tetrahedra, i.e., a pair of doubly 
charged monopoles). 
This is to be compared for instance, with flipping a spin belonging to one 
of the kagome layers in this configuration, which results in the annihilation 
of two adjacent monopoles and corresponding de-excitation of the tetrahedra 
down to their 2in-2out lowest energy state. 
(c) 
Generic low-energy spin ice configuration, where the spins satisfy the local 
2in-2out ice rules. 
(e)
Configuration belonging to the kagome ice phase (KagIce), intermediate 
between SatIce and spin ice, where all the triangular spins are polarised 
in the field direction while the kagome spins arrange themselves in order 
to satisfy the ice rules. 
This requires every triangle in the kagome planes to have exactly one 
spin with a negative projection in the field direction; the positions of these 
anti-aligned spins can be mapped to a dimer covering of the dual honeycomb 
lattice, as illustrated in (f), with a top view down the $[111]$ axis. 
}
\end{figure*}

Four distinct regimes are encountered in spin ice in presence of a field 
along a [111] crystallographic direction, 
as displayed in Fig.~\ref{fig: 111 phase diagram}~\cite{Bramwell2001}. 
In order to understand this phase diagram, it is convenient to divide 
the spin lattice (pyrochlore, or corner-sharing tetrahedral lattice) into 
alternating kagome and triangular layers, as illustrated in 
Fig.~\ref{fig: three phases}. 
Further details on spin ice systems as well as on their phase diagram 
in a $111$ field can be found in the Suppl. Info. 

In the limit of strong fields, in the \textit{saturated} (SatIce) regime, 
all the spins point along the field direction, while respecting the local 
easy axes (see Fig.~\ref{fig: three phases}, top middle panel). 
The ice rules are violated everywhere and each tetrahedron hosts a 
monopole; the monopoles form an ``ionic crystal''. 

As the field strength is reduced, violations of the ice rules are no longer 
sufficiently offset by a gain in Zeeman energy and a regime where most 
tetrahedra obey the ice rule is recovered for $T \ll \Delta$. 
This necessarily requires some of the spins to point against 
the applied field. At intermediate field strengths, they will dominantly be
the spins in the kagome planes, as their Zeeman energy is smaller by a factor
of $3$ compared to the spins in the triangular planes. This leads to 
an extensively degenerate regime known as \textit{kagome ice} (KagIce), 
illustrated in the bottom middle panel in Fig.~\ref{fig: three phases}. 

At low field strengths, the kagome ice regime becomes entropically unstable to 
the `conventional' \textit{spin ice regime}, namely the ensemble of all 
configurations satisfying the ice rules irrespective of the polarisation of 
the triangular spins (top right panel in Fig.~\ref{fig: three phases}). 
All of these regimes crossover at sufficiently large temperatures 
into a conventional \textit{paramagnetic} regime. 

These different regimes appear in the field-temperature phase diagram of 
spin ice as illustrated in Fig.~\ref{fig: 111 phase diagram}. Note that only 
the KagIce and SatIce phases are separated at low temperatures by an actual 
(first order) phase transition~\cite{Hiroi2003,Sakakibara2003} 
(see Suppl. Info. for further details). 
%
%

\subsection{
Field quench protocols and non-equilibrium phenomena
           }
In this article we discuss field quenches from SatIce to 
KagIce, as exemplified by the blue vertical arrows in 
Fig.~\ref{fig: 111 phase diagram}. 
We focus on field quenches at (constant) temperature 
$T \lesssim 0.6\,\textrm{K}\ll \Delta$, as the system appears to relax 
straightforwardly to equilibrium at higher temperatures (blue dash-shaded 
region in Fig.~\ref{fig: 111 phase diagram}) on time scales of the order of 
a typical microscopic spin-flip time scale.

We are interested here mostly in the low-temperature dynamics that 
follows quenches across the first order phase transition in 
Fig.~\ref{fig: 111 phase diagram}. As we shall see in the following, these 
quenches are characterised by interesting slow relaxation phenomena and 
dynamical arrest. By comparison, quenches at higher temperatures 
$T \sim 0.5-0.6\,\textrm{K}$ appear to be fast and featureless. 
The presence of a critical point in the phase diagram at these higher 
temperatures raises a number of separately interesting questions and ideas for 
further work, such as the possibility of quenching to / across a critical 
point, and testing whether the critical correlations bear any signatures in 
the resulting dynamical behaviour. Whilst interesting in their own right, 
these issues are beyond the scope of the present work. 

The dynamics of spin ice at $T \lesssim 5$~K appears to be based on 
incoherent spin reversals, well modelled by a Monte Carlo (MC) 
single-spin flip dynamics\cite{Jaubert2009}. From AC susceptibility 
measurements~\cite{Snyder2004}, a Monte Carlo time step corresponds to 
approximately 1~ms in Dy$_2$Ti$_2$O$_7$, which we use 
to convert MC time into physical time in this work. 

We used Monte Carlo simulations with single spin flip dynamics, with 
{\DTO} spin ice parameters as in Ref.~\cite{Melko2004}. 
We employed Ewald summation techniques~\cite{deLeeuw1980} to treat the long 
range dipolar interactions, and the Waiting Time Method~\cite{WTM_refs} to 
access long simulation times at low temperatures (see also 
Ref.~\cite{Castelnovo2010}). 

\begin{figure*}
\includegraphics[]
                {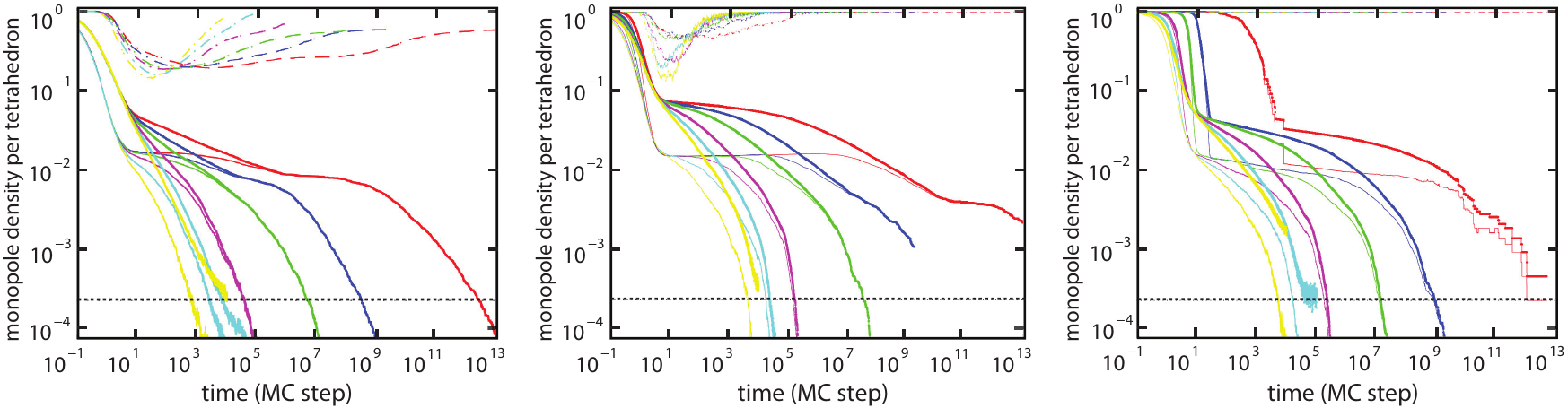}
\caption{\label{fig: overall decay} 
Monopole density (thick lines), density of triangular spins in the direction 
of the initial magnetization (thin dot-dashed lines), and density of 
noncontractible pairs (thin solid lines) from Monte Carlo simulations 
for a system of size $L=8$ ($8192$ spins), 
fields $H=0.2,0.4,0.6$~Tesla (left, middle, 
and right panel, respectively) and temperatures 
$T=0.1,0.15,0.2,0.3,0.4,0.5$~K (red, blue, green, magenta, cyan, yellow). 
Note that the thick lines represent the total monopole density, inclusive 
of those involved in noncontractible pairs; therefore the thick lines are 
always higher or at most equal to the thin lines of the same colour. 
All triangular spins are initially polarised in the direction of the 
the applied field and their initial density is $1$. For sufficiently low 
temperatures with respect to the strength of the applied field, their density 
becomes again approximately one at long times (i.e., in equilibrium). 
At intermediate times, some of the triangular spins reverse, as shown by the 
dip in the density; the latter has been magnified by a factor of 100 and 1000 
in the left and middle panels, respectively, for visualization purposes. 
In the right panel the density of triangular spins in the direction of the 
applied field remains $1$ throughout the simulations, as expected 
for fields  $H > 0.42$~Tesla (see main text and SI). 
In this case the triangular spins remain heavily polarised throughout 
the quench, and the monopole motion is effectively two-dimensional. 
Notice that at the longest times, practically all monopoles are in 
non-contractible pairs. 
(Here and in later figures the black dotted horizontal line indicates the 
density threshold of one monopole in the entire Monte Carlo system.) 
}
\end{figure*}

Observables that we monitor after a field quench are monopole 
density and magnetisation, as illustrated in Fig.~\ref{fig: overall decay}. 
In addition to gaining insight on non-equilibrium dynamical phenomena in 
spin ice -- in particular on the role of the emergent magnetic monopoles -- 
our aim is to investigate whether a protocol can be devised that allows to 
prepare spin ice samples in a low-temperature yet monopole-rich state by 
means of a rapid reduction of an applied $[111]$ magnetic field. 
Attaining such state would be a significant step towards a more direct 
experimental detection of the magnetic monopoles, for instance using 
zero-field NMR techniques~\cite{Sala2012}, 
as we discuss towards the end of the article. 

The use of field quenches instead of thermal quenches~\cite{Castelnovo2010} 
offers two practical advantages: 
(i) a fast variation in the externally applied magnetic field is 
experimentally less challenging than performing a controlled thermal quench; 
and 
(ii) in quenches from SatIce to KagIce, the monopole density is to an 
excellent approximation proportional to the magnetisation of the sample, 
a thermodynamic quantity, and 
therefore directly and simply accessible in experiments. 

In order to make contact with a broader range of experimental settings, 
we briefly review towards the end of the paper alternative experimental 
protocols, such as quenches to zero field, as well as slow ``quenches'' 
(i.e., finite-rate field ramps). 
For completeness, these are discussed in greater detail in the Supporting 
Information. 

We point out that in the following every reference to the 
\textit{magnetic field} is intended as the externally applied field $H$. 
Furthermore, we note that Monte Carlo simulations are devoid of 
demagnetisation corrections, which ought to be taken into account when 
comparing to possible experimental results. 
%
%

\section{
Saturation to kagome ice
        }
As illustrated in Fig.~\ref{fig: three phases}, KagIce comprises the hardcore 
dimer coverings of the honeycomb lattice. The equilibrium density of 
monopoles in this regime is exponentially small at low temperatures, 
and they correspond to monomers in the dimer language. 
SatIce instead corresponds to an ionic crystal of monopoles, i.e., a 
configuration without any dimers. 
Microscopically, in SatIce pairs of 
neighbouring positive and negative monopoles on the same kagome plane 
annihilate by inverting the intervening spin, which is equivalent to the 
deposition of a dimer on the corresponding bond in the honeycomb lattice. 

Field quenches from SatIce to KagIce can thus be understood as a stochastic 
deposition process of dimers, with non-vanishing desorption probability, in 
presence of long range (3D) Coulomb interactions between the monomers 
(i.e., the monopoles). Eventually the dimers become fully packed, up to an 
exponentially small concentration of thermally excited monomers at equilibrium. 

The physics of dimer deposition processes with only \emph{short range} 
interactions was recently considered in the experimental context of molecular 
random tilings~\cite{Blunt2008}, which were shown to exhibit 
a variety of different (stochastic) dynamical regimes akin to kinetically 
constrained models~\cite{Stinchcombe2001,Garrahan2009}. 
In the following we show how the presence of long range interactions, as well 
as additional kinematic constraints in the motion of monopoles due to the 
underlying spin configuration, give rise to remarkably rich behaviour. 
%
%

\subsection{
Short-times: dimer deposition
           }
\begin{figure}
\includegraphics[]
                {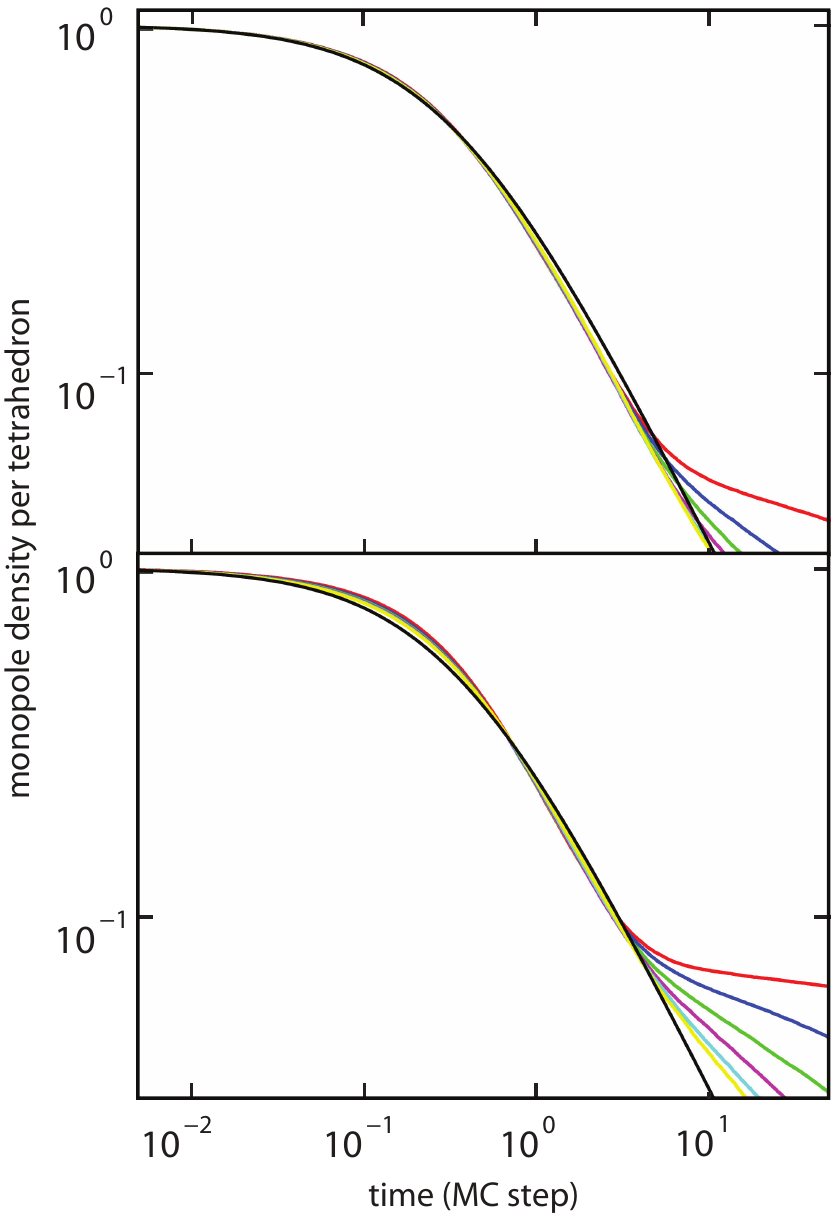}
\caption{\label{fig: init decay} 
Comparison of the initial monopole density decay in simulations with the 
analytical mean field result in Eq.~\eqref{eq: init decay, low field} 
(black solid line). 
The Monte Carlo simulations are for an $L=8$ system, with 
$H=0.2$~Tesla (left panel) and $H=0.3$~Tesla (right panel), 
$T=0.1,0.2,0.3,0.4,0.5,0.6$~K (red, blue, green, 
magenta, cyan, and yellow curves). 
}
\end{figure}

The initial spin flip events following a quench from SatIce to KagIce lower 
the energy of spin interactions by removing monopoles, but they increase the 
Zeeman energy. As the latter is proportional to the quench field, at weak 
fields, all flips are initially downhill in energy and thus take place on a 
time scale of the order of 1 MC step, independently of temperature. 
A mean-field equation capturing the initial decay of the monopole density is
\bea
\frac{d\rho(t)}{dt} &=& - \frac{3}{\tau_0} \rho^2(t) 
\qquad 
\rho(t=0) = 1
\\ 
\rho(t) &=& \frac{1}{1 + 3(t/\tau_0)} 
\label{eq: init decay, low field} 
\eea
where $\rho$ is the monopole density per tetrahedron and the factor 3 
appears because each monopole can equally annihilate with three of 
its neighbours. 
This is in good agreement with the simulation results in 
Fig.~\ref{fig: init decay}. 

For quench fields above a threshold of $0.3$~Tesla, the initial decay becomes 
activated, as the Zeeman energy for reversing a kagome spin 
$20 \mu_B H / 3 \simeq 4.48 H$~K exceeds the spin-spin interaction 
energy gain of about $-1.34$~K ({\DTO} parameters)~\cite{footnote1}. 
For fields beyond this threshold, the initial decay is described by a dimer 
deposition process with field-dependent energy barriers. 

The near-neighbour monopole pair annihilation continues until no suitable 
pairs are left. In dimer deposition language, this corresponds to the random 
packing limit where it is no longer possible to deposit futher dimers without 
first rearranging those already present on the lattice. 

For a wide range of fields and temperatures, the initial decay takes 
approximately $10$~MC steps (see Fig.~\ref{fig: overall decay}) and reaches a 
monopole density in the range 0.05 - 0.08, depending on field and temperature.
 
We compare this value to that obtained from numerical simulations of random 
depositions of dimers on the honeycomb lattice, where we 
measure the density of vacancies at the end of each deposition process when 
it is no longer possible to add a (hard-core) dimer to the system without 
rearranging the ones already deposited. The density of leftover vacancies in 
the non-interacting simulations equals $0.121 \pm 0.003$. 
It is interesting to notice that this value is at least $50$\% larger 
than that following field quenches from SatIce to KagIce. We find that the 
difference is due to the combination of a 
non-vanishing desorption probability and long range Coulomb attraction 
between oppositely charged monopoles, which are present in the spin ice 
quenches and not in the random deposition process. 
%
%

\subsection{
Long times: from noncontractible pairs 
to dynamically-obstructed monopole diffusion
           }
At the end of the initial decay, no neighbouring monopoles are left in any 
of the kagome planes. The majority of monopoles are `isolated' and only a 
minority forms (noncontractible) pairs across adjacent planes 
(\ref{fig: overall decay}). 


Two oppositely-charged monopoles on neighbouring tetrahedra form a 
\emph{noncontractible} pair whenever inverting the direction of the shared 
spin leads to an even higher energy configuration with the two tetrahedra in 
the 4in and 4out state, respectively~\cite{Castelnovo2010}.
For instance, every monopole pair across a triangular spin in SatIce is 
noncontractible (see Fig.~\ref{fig: three phases}). 
Indeed, in KagIce, where the triangular spins are polarised in the field 
direction, noncontractible pairs can only form between positive and negative 
monopoles belonging to \emph{adjacent} kagome planes.

For small values of the applied field ($H \sim 0.2$~Tesla), 
we see from Fig.~\ref{fig: overall decay} that the decay in the density 
of the isolated monopoles is fast in comparison to the decay of 
noncontractible pairs, and eventually only the latter are left to 
determine the long time behaviour of the system. 
This regime can be understood in analogy to thermal 
quenches~\cite{Castelnovo2010} (see Suppl. Info.). 

Whereas isolated monopoles are essentially free to move, noncontractible 
pairs have an activation barrier to move/decay, and thus their lifetime 
increases exponentially with inverse temperature. 
This explains why the isolated monopoles decay faster than noncontractible 
pairs; it also explains the mechanism responsible for the long time decay in 
the overall monopole density of the system. 
Whereas lattice-scale kinematic constraints control the formation (or 
possibly the survival from the initial SatIce state) of noncontractible pairs, 
it is their decay that ultimately controls the long time behaviour of the 
monopole density. 

The life time of a noncontractible pair is determined by the energy barriers 
of the allowed decay processes (namely, processes where the two charges 
separate and annihilate one another elsewhere on the lattice; or they may 
annihilate with other charges in the system). 
For quenches to zero field, the energy barriers are solely due to spin-spin 
interactions and the corresponding decay occurs via a process, 
identified in thermal quenches~\cite{Castelnovo2010}, whereby the 
noncontractible pair recombines by hopping the long way around a hexagon 
containing the obstructing spin. 

The presence of a small applied field does not alter the picture qualitatively, 
but it introduces two important differences. 
Firstly, the mechanism and barrier to the decay of noncontractible 
pairs is altered. Once the majority of the triangular spins are polarised, 
it is straightforward to see that the hexagonal processes invoked in 
Ref.~\onlinecite{Castelnovo2010} are no longer available: decay must proceed 
by separation of the monopoles until they meet oppositely charged ones 
from other noncontractible pairs. 
This process incurs a Coulomb energy barrier for separating 
the positive and negative charges in a noncontractible pair, which grows with 
the distance to the next monopole in the plane. 
Moreover, the presence of an applied magnetic field also affects the 
value of the barrier to separation, as the Zeeman energy for spin flips needs 
to be taken into consideration along with the Coulomb attraction of the 
oppositely charged monopoles. 

Secondly, isolated monopoles hopping within kagome planes exhibit 
activated behaviour of their own due to the Zeeman energy (see 
below for details). 
This explains the opening of an intermediate time window between the end 
of the initial decay, discussed in the previous section, and the final 
decay controlled by noncontractible pairs. 
This time window is absent in thermal quenches~\cite{Castelnovo2010} as well 
as in field quenches down to zero field (see Suppl. Info.). 

This intermediate-time regime is governed by diffusion-annihilation processes 
of isolated monopoles across larger and larger distances as their 
density is reduced. 
Depending on the value of the field and temperature, our simulations 
show different behaviours following the initial decay, including an apparent 
power-law at low fields (see left panel in 
Fig.~\ref{fig: overall decay}). 
This regime is controlled by ``finite time'' processes rather than by any 
asymptotic behaviour. 
Albeit challenging to model in the absence of an asymptotic regime, it is a 
novel and interesting example of a 
reaction-diffusion process in presence of long range Coulomb interactions 
and kinematic constraints that is experimentally accessible in spin ice 
materials. 

As the field strength increases, the decay of the isolated monopoles gets 
comparatively slower, until for $H \sim 0.6$~Tesla they remain the 
majority with respect to noncontractible pairs at all time. In this regime, 
a new mechanism controls the long time behaviour:
the system remains always in a Coulomb liquid phase, rather than condensing 
its monopoles into noncontractible pairs. It is the isolated monopoles 
themselves that, by becoming exceedingly slow, determine the long time decay 
of the system. 

In order to better understand and confirm this scenario, let us look in 
detail at how such slowing down of the monopole time scales takes place. 
Consider the motion of a monopole within a kagome layer (the illustrative 
Fig.~\ref{fig: nn mono hopping} 
provided in the Suppl. Info. may be of help here). 
Ordinary monopole motion involves alternatively flipping pairs of spins with 
negative and positive projection onto the applied field. 
This yields a periodic energy landscape due to the Zeeman 
energy change $\pm 4.48 H$ for the kagome spins, where $H$ is measured 
in Tesla and the energy is measured in degrees Kelvin. 
As the field increases from zero, this barrier progressively slows down the 
monopoles. 

For sufficiently large applied fields, another process becomes energetically 
more convenient: A  new monopole-antimonopole pair is first excited 
neighbouring the existing monopole, which then annihilates with the 
opposite member of the pair (\emph{pair-assisted} hopping -- 
see figure in the Supplementary Material). 
The energy barrier for this process is 
$
\Delta_s - E_{\rm nn} + E_{\rm 2n} - 4.48H 
\simeq 
4.45 - 4.48 H
$, 
where $\Delta_s \simeq 5.64$~K is the cost of flipping a spin in spin ice 
and $E_{\rm nn} - E_{\rm 2n} \simeq 1.19$~K is the Coulomb energy difference 
between two monopoles at nearest-neighbour vs second-neighbour distance 
({\DTO} parameters). 
The new process becomes energetically more favourable than the ordinary 
Zeeman barrier if $H \gtrsim 0.5$~Tesla. 
Notice that the sign of the Zeeman contribution in the new process is 
reversed: if the fields were increased significantly beyond the threshold, 
monopole diffusion would once again become a fast process. 
However, this is forbidden by an 
intervening first order phase transition~\cite{Castelnovo2008} which already 
affects field quenches to $H > 0.7$~Tesla. 

\begin{figure}
\includegraphics[]
                {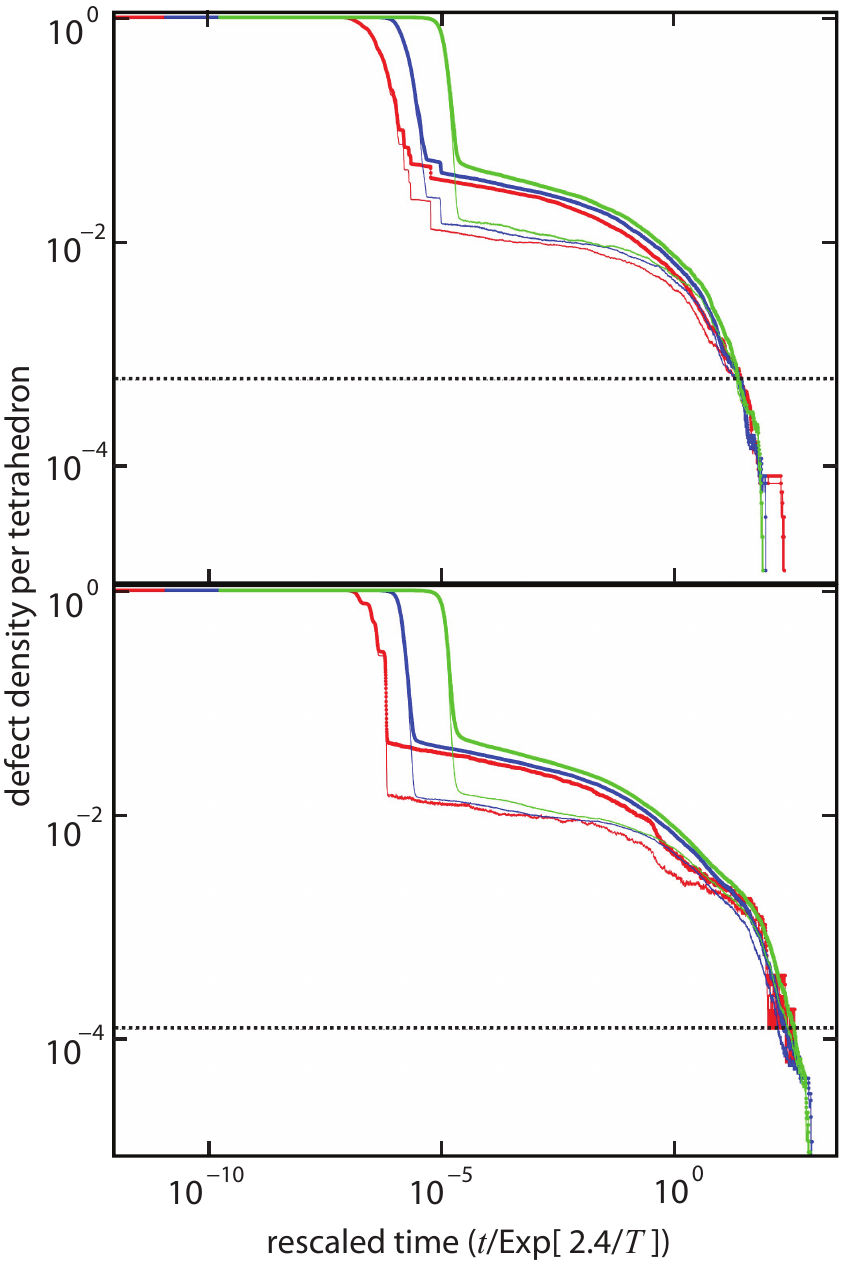}
\caption{\label{fig: long time decay} 
Collapse of the long time decay of the monopole density (thick lines) and 
of the \emph{noncontractible monopole density} (thin lines) 
after rescaling the time axis by a factor $\exp(-2.4 / T)$. 
The Monte Carlo simulations are for $L=6$ and $10$ (left and right panels, 
respectively), with $H=0.6$~Tesla and $T=0.13,0.15,0.18$~K (red, blue, and 
green curves). 
The good quality of the collapse indicates that the simulated systems are 
large enough for the energy scale of $2.4$~K not to exibit appreciable 
system size depedence. 
}
\end{figure}

The largest quench field we could study was $H=0.6$~Tesla. 
At this value, the two processes above give rise to barriers of 
$2.7$~K and $1.8$~K, respectively. These values ought to be corrected 
for and broadened by quadrupolar terms that are missing in the monopole 
picture used in the estimates~\cite{Castelnovo2008}. 
Indeed, the long time decay of the curves in 
Fig.~\ref{fig: overall decay}, collapses upon rescaling the time axis by a 
Boltzmann factor $\exp(- 2.4 / T)$, as illustrated in 
Fig.~\ref{fig: long time decay}. 
The corresponding value of the activation barrier, $2.4$~K, is in reasonable 
agreement with the estimates for the barrier to isolated monopole motion. 

We stress that for $H \gtrsim 0.5$~Tesla, the system enters a long-lived 
Coulomb liquid phase with an enhanced (out-of-equilibrium) density of 
exceedingly slow monopoles. 
This is a regime that may be experimentally promising to probe and detect 
single monopoles, as we discuss farther below. 
%
%

\section{
Experimental probes
        }
We have argued that different quench regimes lead to the appearance 
of a variety of non-equilibrium dynamical phenomena. 
The next question to ask is how best to access such quantities 
experimentally. On the one hand, a solid state system like spin ice does 
not permit direct imaging of the magnetic state in the bulk, unlike 
two-dimensional systems such as artificial spin 
ices~\cite{artificial_spin_ice_refs}. 
On the other hand, we have access to a number of well-developed probes; 
this being a spin system, natural options are magnetisation and NMR. 
%
%

\subsection{
Magnetisation}

One of the advantages of using $[111]$ field quenches from SatIce to KagIce 
is the fact that we have direct access to the monopole density $\rho$ 
per tetrahedron by measuring the magnetisation $M$ in the $[111]$ direction 
(in units of $\mu_B /$~Dy): 
the equation $M(\rho) = (5 \rho + 10)/3$ is extremely accurate for 
$H \gtrsim 0.42$~Tesla, and it remains very useful also at lower 
field strengths, with deviations due to reversed triangular spins of less 
than $1$\% down to $H \sim 0.2$~Tesla (see Supporting Information), 
for $T \lesssim 0.6$~K. 
%
%

\subsection{
Nuclear magnetic resonance}

A promising approach to measure the monopole density in spin ice samples 
follows from the zero-field NMR measurements pioneered by the 
Takigawa group~\cite{TakigawaNMR,Sala2012}. 
This technique uses {\DTO} samples with NMR active  $^{17}$O nuclei at 
the centres of the tetrahedra. The large internal fields induced by the rare 
earth moments at these locations ($\sim 4.5$~Tesla) result in a peak in the 
NMR signal from the $^{17}$O ions in zero external field. 

When a monopole is present in a tetrahedron, the violation of the ice rules 
results in a sizeable reduction of the field at the centre of a tetrahedron 
($\sim 3.5$~Tesla). This in turn ought to give rise to an additional shifted 
peak in the NMR signal. The relative intensity of the two peaks can be 
used for instance to extract the magnetic monopole density in the 
sample~\cite{Sala2012}. 

One of the challenging aspects of the zero-field NMR approach to detect 
monopoles is the trade-off between having a large number of monopoles to 
enhance the signal and having sufficiently long persistence times (the time 
a monopole spends in a given tetrahedron) to reduce the noise in the signal. 
High-temperatures cannot be used to increase the monopole population because 
they lead to fast magnetic fluctuations. On the other hand, the 
system cannot be allowed to thermalise at low temperature as this would result 
in an exponentially small -- thus undetectable -- monopole density. 

The field quenches discussed in this paper can be used to prepare a spin ice 
sample in a strongly-out-of-equilibrium state that offers great potential to 
NMR measurements. Indeed, the long-time regime at low temperatures 
($T < 0.2$~K) has a sizeable density $\rho \sim 0.01$ of 
long-lived, static monopoles (see Fig.~\ref{fig: overall decay}). 

\begin{figure}
\includegraphics[]
                {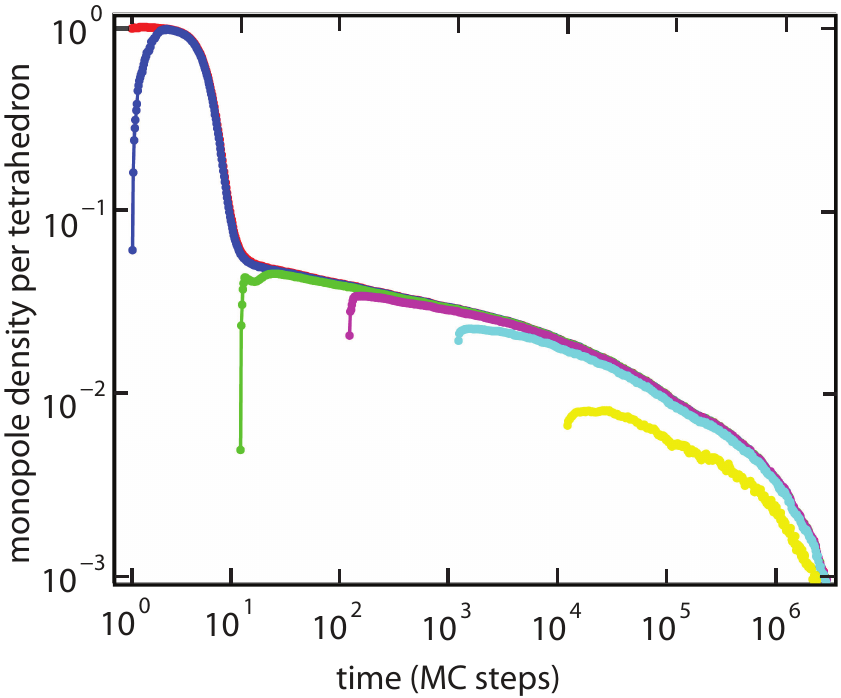}
\caption{\label{fig: rhotimes Hq=06}
Density $\rho(t,\tau)$ of monopoles present in the system at a given 
time $t$, whose persistence time in their current tetrahedron is 
larger than $\tau$. This is shown for a system of size $L=6$ and a quench 
to $H = 0.6$~Tesla at $T = 0.2$~K. 
The different colours correspond to different choices for $\tau$ while the 
curves are presented as a function of $t$ (horizontal axis): 
$\tau = 0.1,1,10,10^2,10^3,10^4$~MC steps 
(red, blue, green, magenta, cyan, and yellow). 
Note that the largest value of the persistence time $\tau = 10^4$~MC steps 
is comparable with the characteristic time scale of the long-time decay 
$\exp(2.4 / T) \sim 10^{5}$~MC steps, which explains the 
slight departure of the yellow curve from the others. 
It is clear that it is thus possible to realise monopole densities of the 
order of $1 \%$ which persist on the scale of minutes! 
} 
\end{figure}

In Fig.~\ref{fig: rhotimes Hq=06} we show the density of monopoles 
$\rho(t,\tau)$ that are present in the system at a given time $t$ 
and that have remained in their current tetrahedron for longer than $\tau$ 
(persistence time). 
Clearly, $\rho \equiv \rho(t,0)$ and $\rho(t,\tau) = 0$, $\forall\,t<\tau$. 
Fig.~\ref{fig: rhotimes Hq=06} confirms that the long time decay of the 
monopole density is indeed due to long persistence times of the residual 
(essentially static) monopoles rather than mobile monopoles which are somehow 
unable to annihilate each other. 
%
%

\subsection{
Alternative experimental protocols
           }
In practice, depending on details of the set-up, other protocols may be easier 
to implement than sudden quenches 
to non zero field values. For instance, it may be more practical to remove a 
field altogether -- quenching to SpinIce rather than to KagIce -- or to 
lower the field gradually to allow time to dissipate the field energies. 
We present here only a brief summary of the results. Details can be found in 
the relevant sections of the Supporting Information. 

We find that zero-field quenches lack the intermediate-time regime between 
the end of the dimer deposition process and the onset of the long time decay 
controlled by noncontractible pairs (this intermediate-time regime is 
evident for instance in Fig.~\ref{fig: overall decay}). 
Once again, one concludes that zero field quenches can be used to prepare 
spin ice samples in a metastable state rich in monopoles forming 
noncontractible pairs, albeit the process appears to be about one order of 
magnitude less efficient than SatIce to KagIce quenches with respect to 
the resulting density. 

Replacing an instantaneous field quench (on the scale of the microscopic 
time $\tau_0 = 1$~MC step $\sim 1$~ms) with a constant-rate field sweep 
clearly has a profound effect on the short time behaviour -- it corresponds to 
an annealed dimer deposition process, which would itself be an interesting topic 
for further study. However, shortly after the field reaches its final value, 
MC simulations show that the evolution of the monopole density agrees 
quantitatively with the behaviour following a field quench to the same field 
value. 

Therefore, any experimental limitations to achieve fast variations of 
applied magnetic fields affect only the behaviour of the system 
while the field is varied, whereas the main features discussed above  
occurring on a longer time scale remain experimentally accessible. 
%
%

\section*{Acknowledgments}
This work was supported in part by EPSRC grants EP/G049394/1 and 
EP/K028960/1 (C.C.). This work was in part supported by the Helmholtz Virtual 
Institute ``New States of Matter and Their Excitations''. C.~Castelnovo 
acknowledges hospitality and travel support from the MPIPKS in Dresden. 
We are grateful to M.~Takigawa and K.~Kitagawa, R. Kremer, S. Zherlytsin and 
to T.~Fennell for discussions on experimental probes of field quenches in 
spin ice.
%
%

\section{
Conclusions
        }
Spin ice offers a realization of several paradigmatic concepts in 
non-equilibrium dynamics: dimer absorption, Coulombic reaction-diffusion 
physics and kinetically constrained slow dynamics. There is an unusually 
high degree of tuneability: the timescale of the elementary dynamical move 
through a Zeeman energy barrier; the dimensionality of the final state 
(d = 2 KagIce vs d = 3 SpinIce); and the relative importance of dimer 
desorption compared to Coulomb interactions. 
The additional availability of a range of experimental 
probes thus allows broad and detailed experimental studies.

With the advent of incipient thermal ensembles of artificial spin ice, where 
the constituent degrees of freedom can even be imaged individually in real 
space, there even appears an entirely new setting for the study of these 
phenomena on the horizon~\cite{thermal_ASI}. 
%
%


\begin{thebibliography}{99}

\bibitem{Stinchcombe2001}
Stinchcombe R (2001) 
Stochastic non-equilibrium systems. 
{\it Adv Phys} 50: 431-496. 

\bibitem{Angell1995}
Angell C A (1995) 
Formation of Glasses from Liquids and Biopolymers. 
{\it Science} 267: 1924-1935. 

\bibitem{Tokuyama2004}
8th Tohwa University International Symposium, Fukuoka, Japan, November 1998. 
{\it Slow Dynamics in Complex Systems}, 
eds Tokuyama M and Oppenheim I (Springer), 
AIP Conference Proceedings, Vol.~469 (2004).

\bibitem{Fredrickson1984}
Fredrickson G H, Andersen H C (1984) 
Kinetic Ising-model of the glass-transition. 
{\it \prl} 53: 1244-1247. 

\bibitem{Ritort2003}
Ritort F, Sollich P (2003) 
Glassy dynamics of kinetically constrained models. 
{\it Adv Phys} 52: 219-342. 

\bibitem{Toussaint1983}
Toussaint D, and Wilczek F (1983) 
Particle-antiparticle annihilation in diffusive motion. 
{\it J Chem Phys} 78: 2642-2647. 

\bibitem{Bramwell2001}
Bramwell S T and Gingras M J P (2001) 
Spin ice state in frustrated magnetic pyrochlore materials. 
{\it Science} 294: 1495-1501. 

\bibitem{Castelnovo2008}
Castelnovo C, Moessner R, and Sondhi S L (2008) 
Magnetic monopoles in spin ice. 
{\it Nature} 451: 42-45.

\bibitem{Castelnovo2012}
Castelnovo C, Moessner R, and Sondhi S L (2012) 
Spin Ice, Fractionalization, and Topological Order. 
{\it Annu Rev Condens Matter Phys} 3: 35-55.

\bibitem{Castelnovo2010}
Castelnovo C, Moessner R, and Sondhi S L (2010) 
Thermal quenches in spin ice. 
{\it \prl} 104: 107201-4.

\bibitem{Levis2012}
Levis D and Cugliandolo L F (2012) 
Out of equilibrium dynamics in the bidimensional spin-ice model. 
{\it EPL} 97: 30002-6; 
Levis D and Cugliandolo L F (2013) 
Defects dynamics following thermal quenches in square spin-ice. 
{\it \prb} 87: 214302-14. 

\bibitem{Ryzhkin2005}
Ryzhkin I A (2005) 
Magnetic Relaxation in Rare-Earth Oxide Pyrochlores. 
{\it JETP} 101: 481-486. 

\bibitem{Jaubert2009}
Jaubert L D C and Holdsworth P C W (2009) 
Signature of magnetic monopole and Dirac string dynamics in spin ice. 
{\it Nat Phys} 5: 258-261.

\bibitem{Matsuhira2000}
Matsuhira K, Hinatsu Y, Tenya K, and Sakakibara T (2000) 
Low temperature magnetic properties of frustrated pyrochlore ferromagnets 
Ho$_2$Sn$_2$O$_7$ and Ho$_2$Ti$_2$O$_7$. 
{\it J Phys: Condens Matter} 12: L649-656. 

\bibitem{Snyder2004}
Snyder J et al. (2004) 
Low-temperature spin freezing in the Dy$_2$Ti$_2$O$_7$ spin ice. 
{\it \prb} 69: 064414-6.

\bibitem{Henley2013}
Henley C L (2013) 
NMR relaxation in spin ice due to diffusing emergent monopoles. 
{\it arXiv:1210.8137}. 

\bibitem{Matsuhira2011}
Matsuhira K, et al. (2011) 
Spin Dynamics at Very Low Temperature in Spin Ice Dy$_2$Ti$_2$O$_7$. 
{\it J Phys Soc Japan} 80: 123711-4

\bibitem{Yaraskavitch2012}
Yaraskavitch L R, et al. (2012) 
Spin dynamics in the frozen state of the dipolar spin ice material 
Dy$_2$Ti$_2$O$_7$. 
{\it \prb} 85: 020410(R)-5. 

\bibitem{Revell2012}
Revell H M, et al. (2012) 
Evidence of impurity and boundary effects on magnetic monopole dynamics 
in spin ice. 
{\it Nat Phys} 9: 34-47. 

\bibitem{Henley2010}
Henley C L (2010) 
The Coulomb phase in frustrated systems. 
{\it Ann Rev Condens Matter Phys} 1: 179-210. 

\bibitem{Hiroi2003}
Hiroi Z, Matsuhira K, Takagi S, Tayama T, and Sakakibara T (2003) 
Specific heat of kagome ice in the pyrochlore oxide Dy$_2$Ti$_2$O$_7$. 
{\it J Phys Soc Japan} 72: 411-418. 

\bibitem{Sakakibara2003}
Sakakibara T, Tayama T, Hiroi Z, Matsuhira K, and Takagi S (2003) 
Observation of a liquid-gas-type transition in the pyrochlore spin ice 
compound Dy$_2$Ti$_2$O$_7$ in a magnetic field. 
{\it \prl} 90: 207205-4. 

\bibitem{Melko2004}
Melko R G and Gingras M J P (2004) 
Monte Carlo studies of the dipolar spin ice model. 
{\it J~Phys:~Condens~Matter} 16: R1277-R1319. 

\bibitem{deLeeuw1980}
de Leeuw S W, Perram J W, and Smith E R (1980) 
Simulation of electrostatic systems in periodic boundary conditions. 
I: Lattice sums and simulation results. 
{\it Proc~R~Soc~London~Ser~A} 373: 27-56. 

\bibitem{WTM_refs}
Dall J and Sibani P (2001) 
Faster Monte Carlo simulations at low temperatures. The waiting time method. 
{\it Computer~Phys~Comm} 141: 260-267. 

\bibitem{Sala2012}
Sala G, et al. (2012) 
Magnetic Coulomb Fields of Monopoles in Spin Ice and Their Signatures 
in the Internal Field Distribution. 
{\it \prl} 108: 217203-4. 

\bibitem{Blunt2008}
Blunt M O, et al. (2008) 
Random tiling and topological defects in a two-dimensional molecular network. 
{\it Science} 322: 1077-1081.

\bibitem{Garrahan2009}
Garrahan J P, Stannard A, Blunt M O, and Beton P H (2009) 
Molecular random tilings as glasses. 
{\it Proc Natl Acad Sci USA} 106: 15209-15213.


\bibitem{artificial_spin_ice_refs}
For reviews, see 
Heyderman L J and Stamps R L (2013) 
Artificial ferroic systems: novel functionality from structure, 
interactions and dynamics. 
{\it J Phys: Condens Matter} 25, 363201; 
Nisoli C, Moessner R, and Schiffer P (2013) 
Artificial spin ice: Designing and imaging magnetic frustration. 
{\it Rev Mod Phys} 85, 1473-1490. 

\bibitem{TakigawaNMR}
Kitagawa K and Takigawa M, 
poster at the ESF-HFM meeting 
``\textit{Topics In the Frustration of Pyrochlore Magnets}'' 
in Abingdon, UK (2009). 

\bibitem{thermal_ASI}
Porro J M, Bedoya-Pinto A, Berger A, and Vavassori P (2013) 
Exploring thermally induced states in square artificial spin-ice arrays. 
{\it  New J Phys} 15: 055012-12; 
Zhang S, et al. (2013) 
Crystallites of magnetic charges in artificial spin ice. 
{\it Nature} 500: 553-557; 
Farhan A, et al. (2013) 
Exploring hyper-cubic energy landscapes in thermally active finite 
artificial spin-ice systems. 
{\it Nat Phys} 9: 375-382; 
Farhan A, et al. (2013) 
Direct Observation of Thermal Relaxation in Artificial Spin Ice. 
{\it \prl} 111: 057204-5. 

\bibitem{footnote1}
The energy gain 
of $-1.34$~K for reversing a kagome spin in SatIce was obtained by 
extrapolating MC values to infinite system size. It is in reasonable 
agreement with the value $(2\alpha-1)E_{\rm nn} - 2 \Delta \simeq -1.73$~K 
that obtains from the effective description in terms of magnetic monopoles, 
where $\alpha = 1.64$ is the Madelung constant of the Zincblende structure 
(an ionic diamond lattice), $\Delta \simeq 4.35$~K is the energy cost of a 
monopole and $E_{\rm nn} \simeq 3.06$~K is the energy of two neighbouring 
monopoles ({\DTO} parameters).

\end{thebibliography}

\begin{thebibliography}{99}


\bibitem{Bramwell2001}
Bramwell S T and Gingras M J P (2001) 
Spin ice state in frustrated magnetic pyrochlore materials. 
{\it Science} 294: 1495-1501. 

\bibitem{Castelnovo2012}
Castelnovo C, Moessner R, and Sondhi S L (2012) 
Spin Ice, Fractionalization, and Topological Order. 
{\it Annu Rev Condens Matter Phys} 3: 35-55.

\bibitem{Shastry1999}
Siddarthan R, et al. (1999) 
Ising Pyrochlore Magnets: Low-Temperature Properties, ``Ice Rules'', 
and beyond. 
{\it \prl} 83: 1854-1857. 

\bibitem{Isakov2005}
Isakov S V, Moessner R, and Sondhi S L (2005) 
Why Spin Ice Obeys the Ice Rules. 
{\it \prl} 95: 217201-4. 

\bibitem{Castelnovo2008}
Castelnovo C, Moessner R, and Sondhi S L (2008) 
Magnetic monopoles in spin ice. 
{\it Nature} 451: 42-45.

\bibitem{Pomaranski2013}
Pomaranski D, et al. (2013) 
Absence of Pauling's residual entropy in thermally equilibrated {\DTO}. 
{\it Nat Phys} 9: 353-356. 

\bibitem{Kimura2013}
Kimura K, et al. (2013) 
Quantum fluctuations in spin-ice-like Pr$_2$Zr$_2$O$_7$. 
{\it Nat Comm} 4: 1934-6. 

\bibitem{Castelnovo2010}
Castelnovo C, Moessner R, and Sondhi S L (2010) 
Thermal quenches in spin ice. 
{\it \prl} 104: 107201-4.

\bibitem{denHertog2000}
den Hertog B C and Gingras M J P (2000) 
Dipolar interactions and origin of spin ice in ising pyrochlore magnets. 
{\it Phys~Rev~Lett} 84, 3430-3433.

\bibitem{Castelnovo2011}
Castelnvo C, Moessner R, Sondhi S L (2011) 
Debye-H\"{u}ckel theory for spin ice at low temperature. 
{\it \prb} 84: 144435-14

\end{thebibliography}




%
%


\clearpage
\setcounter{figure}{0}
\makeatletter 
\renewcommand{\thefigure}{S\@arabic\c@figure} 

\begin{center}
{\huge Supplementary Information}
\end{center}

\section{
Dipolar spin ice model
        }
We present here a brief summary of the model and properties of spin ice. 
For a more thorough review, see for instance 
Ref.~\onlinecite{Bramwell2001} and Ref.~\onlinecite{Castelnovo2012}. 

\begin{figure}[h]
\hspace{-0.3 cm}\includegraphics[width=0.45\textwidth]{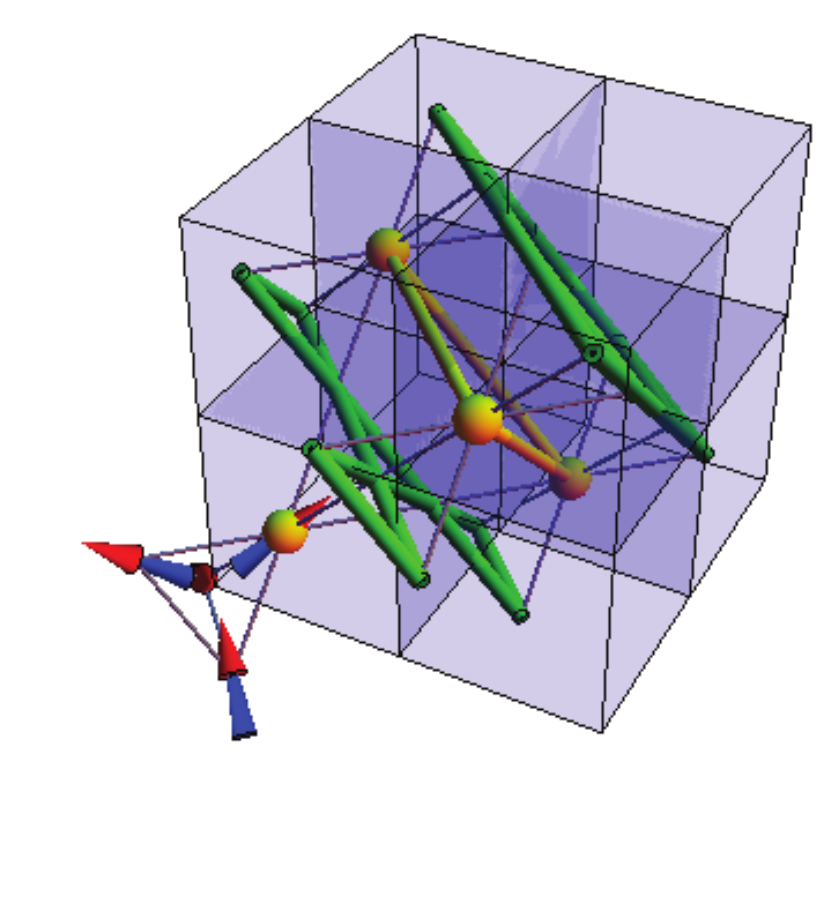}
\caption{\label{fig: tet lattice} 
Illustration of pyrochlore lattice of a spin ice system (the spins are drawn 
only in the bottom left tetrahedron for convenience). 
Given a choice of $[111]$ direction (i.e., one of the major diagonals of 
the cube), the pyrochlore lattice can be seen as a layered structure of 
alternating triangular (yellow) and kagome (green) planes of spins. 
}
\end{figure}

Spin ice models and materials are systems where the magnetic degrees of 
freedom are localised on a lattice of corner sharing tetrahedra 
(pyrochlore lattice), illustrated in Fig.~\ref{fig: tet lattice}. 
Given a choice of $[111]$ direction (i.e., one of the major diagonals of 
the cube), the pyrochlore lattice can be seen as a layered structure of 
alternating triangular and kagome planes of spins. 
This is particularly important in the present manuscript where we will study 
the behaviour of the system in presence of a magnetic field applied in the 
$[111]$ direction, which results in a crucial difference between the 
corresponding triangular and kagome spins. 

\begin{figure}
\hspace{-0.2 cm}\includegraphics[width=0.45\textwidth]{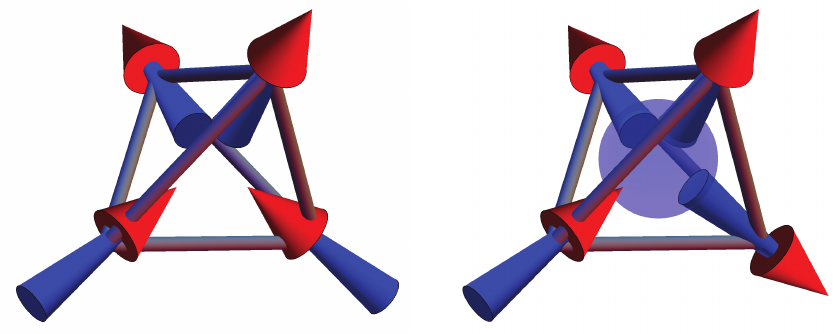}
\caption{\label{fig: single tet} 
Illustration of a spin ice tetrahedron, demonstrating the 4 inequivalent local 
easy axes. The configuration in the left panel corresponds to one 
of the lowest energy (2in-2out) states of the Hamiltonian in 
Eq.~\eqref{eq: Ham}. 
The configuration in the right panel corresponds to a monopole (3in-1out or, 
in this case, 3out-1in); this is the lowest energy excitation in the system. 
Higher energy excitations are obtained by arranging all four spins pointing 
into or out of a tetrahedron. 
}
\end{figure}

The largest energy scale in spin ice systems is a single ion anisotropy that 
forces the spins to lie along the axis connecting the centre of a tetrahedron 
to the corresponding vertex, as illustrated in Fig.~\ref{fig: single tet}. 
This energy scale is usually so large compared to interactions as 
well as the relevant temperature and field energy scales that one can 
model the spins as strictly easy-axis (Ising). 

The spins interact via exchange and dipolar coupling, which in these systems 
happen to be of approximately the same magnitude at nearest-neighbour distance. 
The model used in this paper is known as the dipolar spin ice 
model~\cite{Shastry1999}. 
If we label the sites $i$ of the three-dimensional pyrochlore lattice, 
and the local easy-axis $\hat{e}_i$, each spin can be represented as 
a classical vector $\vec{\mu}_i = \mu \, S_i \, \hat{e}_i$, 
where $\mu$ is the size of the magnetic moment and $S_i = \pm 1$. 

The thermodynamic properties of spin ice are described with good accuracy by 
a Hamiltonian energy that encompasses a uniform nearest-neighbour exchange 
term of strength $J \sim 1-3$~K and long-range dipolar interactions, 
\bea
H \!=\! 
\frac{J}{3} 
\sum_{\langle i j \rangle} S_i S_j 
+ \! 
\frac{D r_{\rm nn}^3}{2} \!
\sum_{i j} 
\left[
  \frac{\hat{e}_i \cdot \hat{e}_j}{\vert r_{ij} \vert^3} 
	- 
	\frac{3 \left( \hat{e}_i\cdot r_{ij} \right)\left( \hat{e}_i\cdot r_{ij} \right)}
	     {\vert r_{ij} \vert^5}
\right]
\label{eq: Ham}
\eea
where $r_{\rm nn} \sim 3.5$~{\AA} is the pyrocholore lattice constant, and 
$D = \mu_0 \, \mu^2 / (4 \pi r_{\rm nn}^3) \sim 1.41$~K. 

Due to a peculiar interplay between interactions, lattice geometry and 
local easy-axis anisotropy, it is not possible to minimise simultaneously 
each term in the Hamiltonian and the system is frustrated. 
To a first approximation, the energy is minimised by configurations 
where each tetrahedron has two spins pointing in and two pointing out 
(2in-2out, Fig.~\ref{fig: single tet})~\cite{Isakov2005}. 
These configuration are in one-to-one correspondence with proton disorder in 
(cubic) water ice, hence the name \textit{spin ice} and the fact that the 
2in-2out rules are referred to as \textit{ice rules}. 

There are many ways to fulfill the ice rule condition on the pyrochlore 
lattice, which result in an extensive degeneracy. 
The ensamble of these configurations is neither ordered nor trivially 
disordered; it is perhaps best understood as an instance of classical 
topological order, whereby the system develops an additional (gauge) 
symmetry at low temperatures~\cite{Castelnovo2012}. 

Excitations over these low energy states 
take the form of defective tetrahedra with one spin pointing in and three out, 
or vice versa (3out-1in, Fig.~\ref{fig: single tet}). 
In recent years it was shown that these defects behave as 
Coulomb-interacting point like \textit{magnetic charges} 
(i.e., a Coulomb liquid) free to move in three 
dimensions~\cite{Castelnovo2008}. 

At closer inspection, the 2in-2out configurations are not exactly degenerate 
(in presence of dipolar interactions). 
Their splitting is however much smaller than the strength of the interactions 
$J,\,D$, and a conventional ordering transition is not observed in the system 
down to a correspondingly small temperature 
(only numerical evidence of this transition is available to date, though see 
Ref.~\onlinecite{Pomaranski2013} for some preliminary experimental results). 
Therefore, these systems exhibit a relatively broad temperature range between 
the cost of a monopole excitation, $T \sim J,D$, down to the transition 
temperature $T_c \ll J,D$, in which their behaviour is captured at a coarse 
grained level by a Coulomb liquid of magnetic charges. 
This description has indeed gone a long way in allowing us to understand the 
behaviour observed experimentally in spin ice materials. 
%
%

\section{
Phase diagram in a $[111]$ magnetic field
        }
The effect of an external magnetic field applied on a spin ice system along 
one of the $[111]$ axes is not the same on all spin sublattices. As it is 
evident from Fig.~\ref{fig: tet lattice} and Fig.~\ref{fig: single tet}, 
the field is aligned with the local easy axis of one spin per tetrahedron 
and canted with respect to the other three spins, which thus incur a lesser 
Zeeman coupling. 
In the alternating kagome-triangular layer interpretation of the pyrochlore 
lattice, the parallel spins live on the triangular planes, whereas the 
canted spins live on the kagome planes (see Fig.~\ref{fig: tet lattice}). 

When the field energy is stronger than spin-spin interactions, the lowest 
energy state where each spin has a positive component in the direction of 
the field has each tetrahedron either in a 1in-3out or in a 3in-1out 
configuration (an example of such configuration is shown in the right panel 
of Fig.~\ref{fig: single tet}). 

This configuration corresponds to a perfect ionic crystal of magnetic charges 
living at the centres of the tetrahedra (which form a bipartite diamond 
lattice). 

As the field is reduced, the interactions favour the restoration of the 
2in-2out ice rules. However, since one out of four spins is more strongly 
Zeeman coupled to the field, the ice rules will be restored predominantly 
via re-arrangement of the canted spins. Namely, the spins in the triangular 
planes remain essentially fully polarised whereas one every three spins 
in the kagome layer gets reversed to point opposite to the field direction 
(left panel in Fig.~\ref{fig: single tet}). 

These new configurations that occur at intermediate field values are not as 
disordered as generic 2in-2out spin ice, yet they are only partially 
polarised and they remain extensively degenerate. They are referred to as 
kagome ice states. These states are, up to thermal fluctuations, devoid of 
monopole excitations. 
\begin{figure*}
\hspace{-0.0 cm}\includegraphics[width=0.95\textwidth]{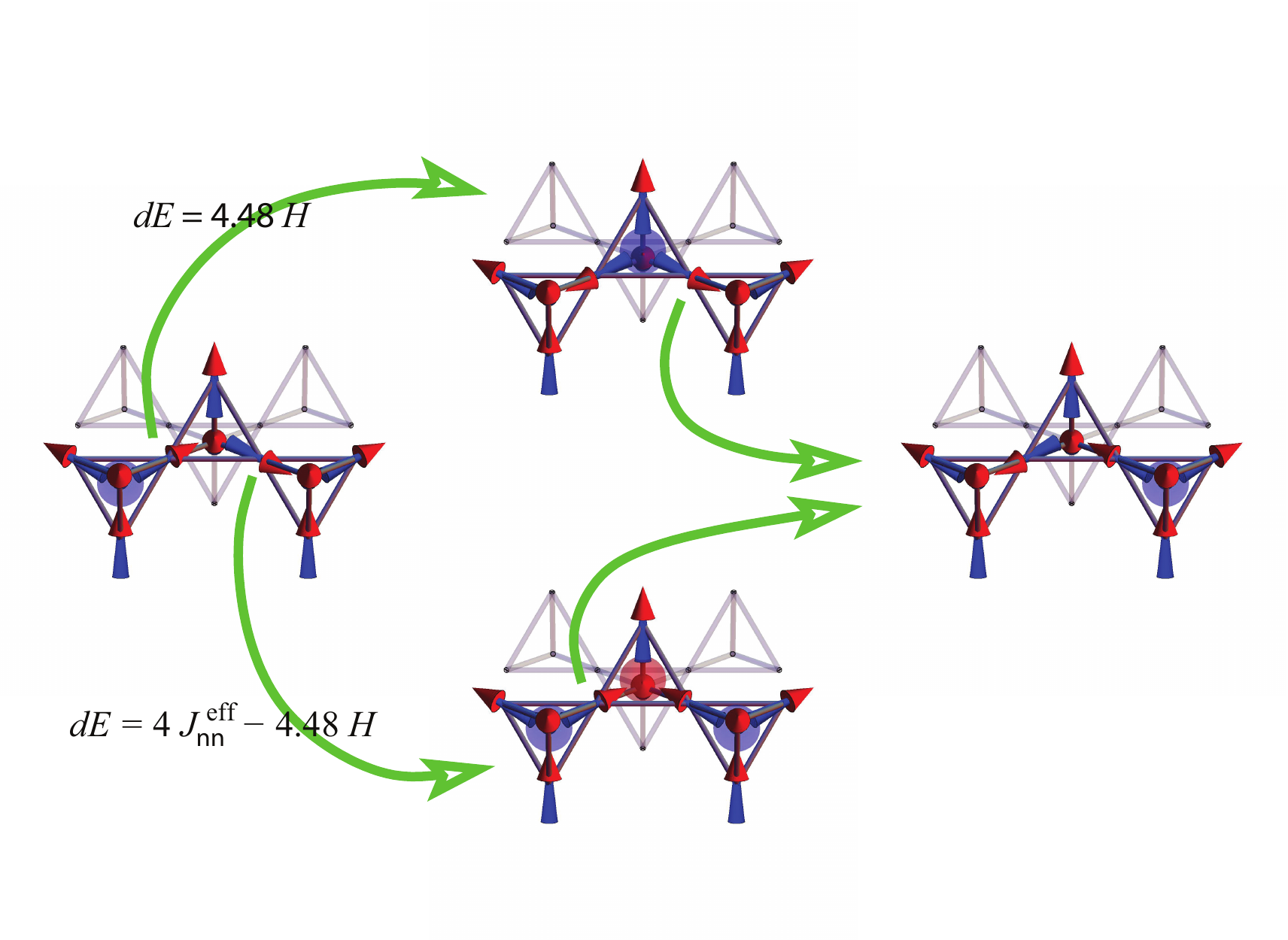}
\caption{\label{fig: nn mono hopping} 
Schematic representation of monopole motion in a kagome plane, both via 
ordinary hopping (top) and via pair-assisted hopping (bottom).
Both processes result in a negative monopole being transferred from a  
downward-pointing tetrahedron (left) to one of the four nearest 
downward-pointing tetrahedra (right). 
The same holds for a positive monopole in an upward-pointing tetrahedron. 
The two processes encompass two spins being flipped each, and 
face different energy barriers $dE$ with opposite field dependence on the 
first spin flip. 
The figure shows the value of the barriers in the nearest-neighbour spin ice 
model. In the main text we discuss how they are modified in presence of 
long-range dipolar interactions; the field dependence however remains 
unchanged. 
The second spin flip that completes the transfer always decreases the 
energy of the system. 
The tails of the green arrows indicate the spin flipped in going from one 
panel to the next. 
Only the spins in the front three tetrahedra are drawn for convenience. 
The triangular spins remain polarised throughout. 
}
\end{figure*}

It is interesting to re-interpret the role of the $[111]$ applied field 
in terms of the Coulomb liquid picture of the excitations of the 
spin ice ground states. From vanishing to intermediate fields, the 
system is able to respond without violating the ice rules, and a continuous 
crossover is observed from spin ice to kagome ice. 
When the field is further increased, it begins to compete with the ice rules, 
as it favours an ionic crystal of magnetic charged. Namely, the applied 
field in the spin language translates into a staggered chemical potential 
in the monopole picture. 

The phase diagram of a liquid of Coulomb interacting charges as a function 
of temperature and staggered chemical potential is the well-known 
liquid-gas phase diagram. This is indeed a rather distinctive phase diagram, 
characteristic of itinerant systems and long range interactions, which is 
quite at odds (unprecedented!) with the behaviour of a localised 
spin system. 
The experimental observation of such phase diagram in spin ice materials 
was indeed one of the first smoking guns that the theoretical proposal 
for emergent magnetic monopole excitations in these systems was indeed 
correct~\cite{Castelnovo2008}. 

Details about this phase diagram can be found in 
Ref.~\onlinecite{Castelnovo2008} (Fig.~4) and it is schematically 
represented in 
Fig.~1 
in the main text. 
At low temperatures, the high-field polarised state (saturated ice) is 
separated from the intermediate field kagome ice state by a first order 
transition. 
In the monopole language, this transition connects a low-density (kagome) 
to a high-density (saturated) monopole phase. In the spin language, the 
order parameter describing the transition is the magnetisation, which 
jumps discontinuously from a low to a high moment. As discussed in the main 
text, up to thermal fluctuations in the triangular spins (which remain 
essentially polarised throughout the transition), the monopole density and 
the $[111]$ magnetisation of the system are directly related 
($M = (5 \rho + 10)/3$). 

At some finite temperature, the line of first order points terminates 
at a characteristic critical end point (Coulombic criticality), and at 
higher temperatures the transition is replaced by a continuous crossover 
from kagome to saturated ice. 
%
%

\section{
Field quenches in nearest-neighbour spin ice
        }
In view of the advent of ``exchange spin ice'' materials, where the Coulomb 
interactions between monopoles is relatively weak~\cite{Kimura2013} 
on account of the super exchange between neighbouring spins dominating over 
their dipolar interactions, it is instructive to 
study the long time behaviour of the monopole density in field quenches 
in the simpler context of nearest-neighbour spin ice. 
There, the lack of dipolar (or further range exchange) interactions implies 
that the 
monopoles are trivially deconfined and no metastable bound states can form. 
For field quenches down to zero field, this removes all possible activation 
barriers and the monopole density is controlled by the asymptotics of 
reaction diffusion processes (see Ref.~\cite{Castelnovo2010}). 

The situation is different in field quenches to finite fields, where there 
exist Zeeman energy barriers to 
monopole motion which thus affect the long time decay of the mononopole 
density. In the discussion hereafter we shall assume for convenience that the 
triangular spins in the system are fully polarized, which is indeed the 
relevant scenario for the long time decay of the monopole density. 

Straightforward hopping of a monopole is object to an alternating energy 
landscape due to the Zeeman energy to flip a kagome spin, $4.48 H$ 
(see main text and Fig.~\ref{fig: nn mono hopping}). 
As this energy scale grows with increasing field, an 
alternative process becomes energetically preferable: the creation of a new 
monopole pair neighbouring the existing monopole, with subsequent 
annihilation of the latter with its opposite member of the pair. 
The energy scale for this process in nearest-neighbour spin ice is 
$4 J^{\rm eff}_{\rm nn} - 4.48H$, which decreases with increasing field. 
The two processes are illustrated schematically in 
Fig.~\ref{fig: nn mono hopping}. 

\begin{figure}
\hspace{-0.0 cm}\includegraphics[width=0.4\textwidth]{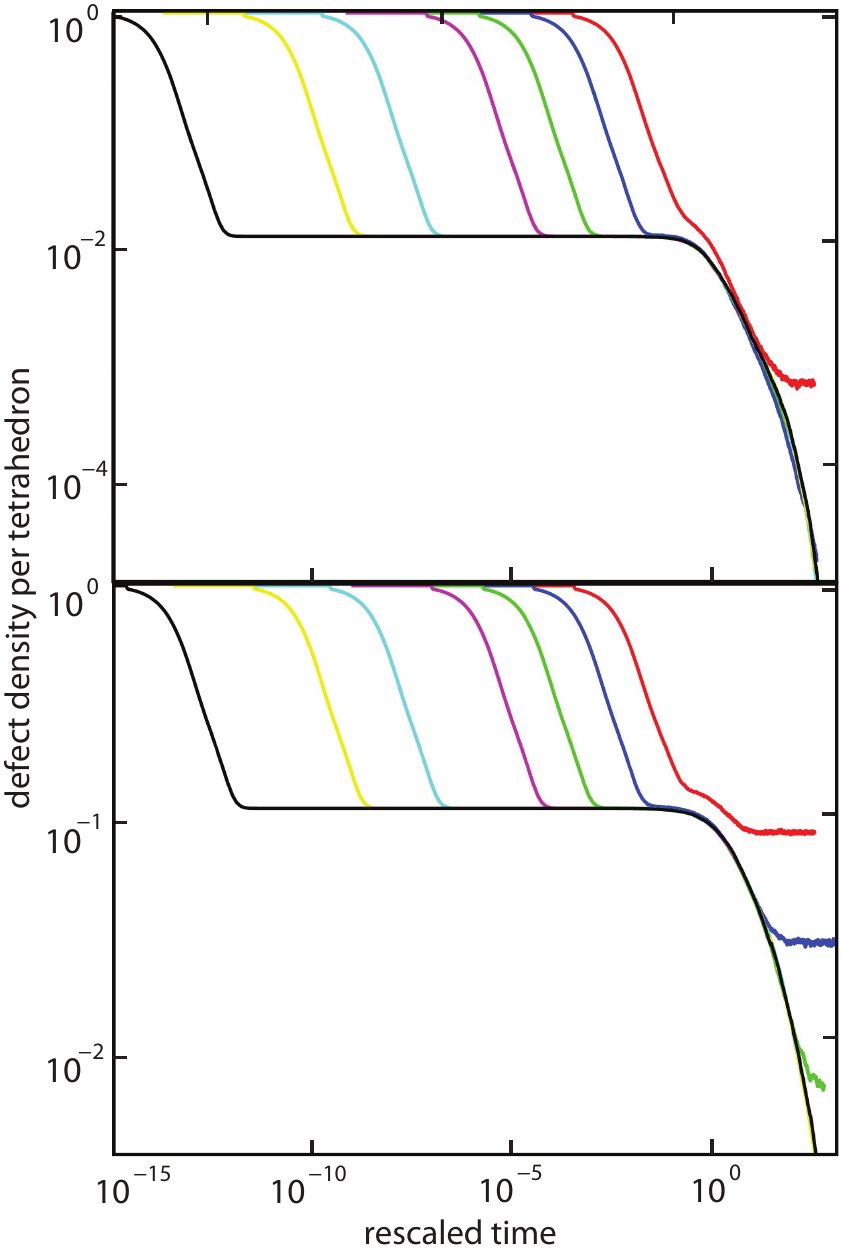}
\caption{\label{fig: nn collapse} 
Density of monopoles for different fields and temperatures from the numerical 
simulations of nearest neighbour spin ice with $J^{\rm eff}_{\rm nn}=1.11$~K 
and $L=6$. 
The time axis has been rescaled by a factor $\exp[-(4.48H)/T]$, 
for $H = 0.4$~Tesla (top panel); and by a factor 
$\exp[-(4 J^{\rm eff}_{\rm nn} - 4.48H)/T]$, for $H = 0.6$~Tesla (bottom panel). 
The different colour curves correspond to different values of the temperature, 
$T=0.5,0.3,0.2,0.15,0.1,0.08,0.06$~K (red, blue, green, magenta, cyan, 
yellow and black). Note the excellent collapse of the final long-time decay 
over approximately 15 orders of magnitude!
}
\end{figure}

For the typical value of $J^{\rm eff}_{\rm nn}=1.11$~K~\cite{denHertog2000}, 
the two energy scales cross over at $H_{\rm th} \simeq 0.5$~Tesla. 
For $H < H_{\rm th}$ we expect that the long time decay of the monopole 
density is controlled by the energy barrier $4.48 H$; for $H > H_{\rm th}$ 
we expect that the relevant barrier is instead 
$4 J^{\rm eff}_{\rm nn} - 4.48H$. 
Fig.~\ref{fig: nn collapse} confirms that the corresponding Boltzmann 
factors lead to a very good collapse of the long time behaviour 
of the monopole density in nearest-neighbour spin ice. 
Note that the very good collapse of the data from nearest-neighbour 
simulations suggests that the \emph{entropic} long range Coulomb interactions 
that are present even in nearest-neighbour spin ice do not play 
a significant role in the long time behaviour of the monopole density at 
these temperatures. 


The introduction of dipolar interactions in the system changes 
the pair-assisted energy scale (even for an isolated monopole), due to the 
Coulomb interaction between the three charges in the intermediate stage. 
In addition to the energy cost of creating two new monopoles, we need to take 
into account two Coulomb interaction terms between opposite charges at 
nearest-neighbour distance and one term between like charges at second 
neighbour distance. 
An estimate of the resulting barrier for typical {\DTO} parameters 
is given in the main text. 
The dependence on the magnetic field by contrast remains unchanged. 
%
%

\section{
The triangular spins
        }
\begin{figure*}
\hspace{-0.1 cm}\includegraphics[width=0.95\textwidth]{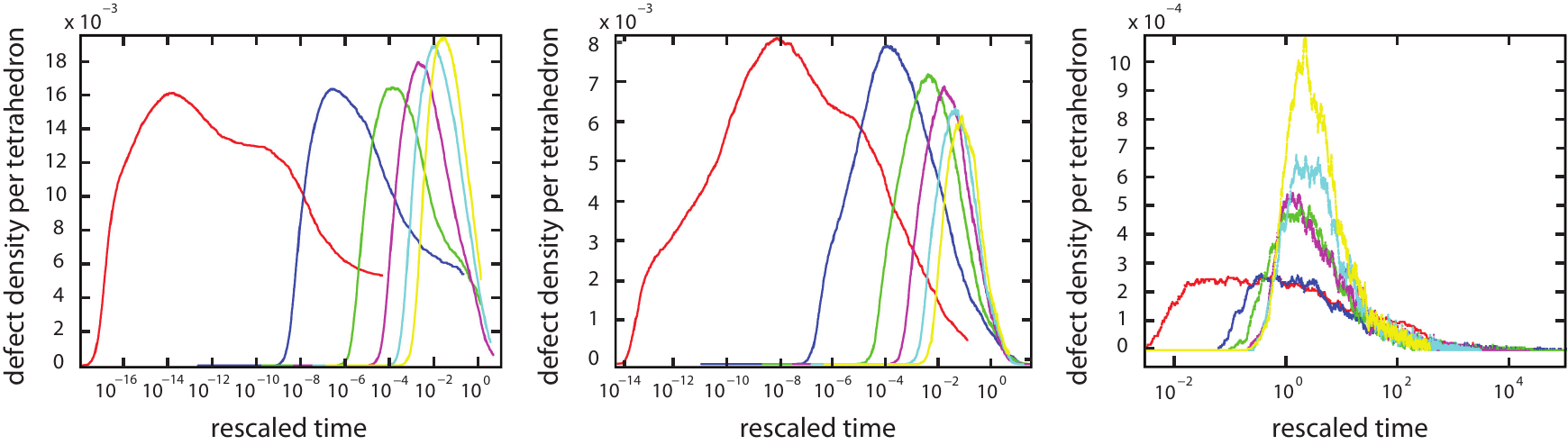}
\caption{\label{fig: tri spins} 
Density of reversed triangular spins for different fields and temperatures 
from the numerical simulations. 
The Monte Carlo simulations are for an $L=8$ system, with 
$H=0.2,\,0.3,\,0.4$~Tesla (left, middle and right panel, respectively), 
$T=0.1,0.2,0.3,0.4,0.5,0.6$~K (red, blue, green, magenta, cyan, and yellow 
curves). 
The time axis in each panel has been rescaled by a 
factor $\exp[-dE/T]$, with $dE = 4.0,\,3.2,\,0.62$~K, from left to right. 
}
\end{figure*}

Much of our interpretation of the behaviour of spin ice following a field 
quench from SatIce to KagIce in this paper relies on the triangular spins 
being almost fully polarised. 
We have verified this assumption explicitly in our numerical simulations 
(see 
Fig.~3 in the main text 
and Fig.~\ref{fig: tri spins}). 
Even at low fields, only a small fraction of triangular spins is
temporarily reversed during the out of equilibrium relaxation. 
In all regimes considered in this manuscript (with the exception of quenches 
down to zero field, to be discussed next), to within $1$\% the magnetisation 
tracks the monopole density as $M(\rho) = (5 \rho + 10) / 3$, in units of 
$\mu_B /$~Dy: the magnetisation is an excellent proxy for the monopole density.

In the following, we discuss what can be learnt from the behaviour of the 
triangular planes alone. 
First of all, for fields $H \gtrsim 0.42$~Tesla, no reversed triangular spin 
can be present in the system for any length of time ($\gtrsim 1$~MC step), as 
it is unstable to flipping and creating two monopoles: the Zeeman splitting 
of a triangular spin $20 \mu_B H / k_B \simeq 13.44\,H$~K exceeds the energy 
cost of monopole pair creation, $\Delta_s \simeq 5.64$~K for {\DTO}. 

When $H \lesssim 0.42$~Tesla, reversed triangular spins can be in a 
(long-lived) metastable state. Our simulations show that this occurs 
mostly towards the end of the initial decay and throughout the intermediate 
decay (see 
Fig.~3); 
the densities 
remain negligibly small even at very low fields: less than $1$\% of the 
triangular spins are reversed at any given time during, say, an 
$H = 0.3$~Tesla  field quench (see Fig.~\ref{fig: tri spins}). 

For $0.42 \gtrsim H \gtrsim 0.32$~Tesla, while it costs energy to flip a 
reversed triangular spin and create two monopoles, the system can then 
readily lower its energy by moving the two monopoles one step each. 
In this field range, the Coulomb cost to separate the monopoles is offset by 
the Zeeman energy gain so that the final state has lower energy than the 
starting one before we flipped the triangular spin. 
This process incurs an energy barrier of approximately 
$\Delta_s - 20 \mu_B H / k_B \simeq 5.64 - 13.44\,H \simeq 0.3$~K at 
$H = 0.4$~Tesla (estimated in the Coulomb liquid approximation), which is in 
reasonable agreement with the value $0.62$~K obtained by collapsing the long 
time behaviour in the right panel of Fig.~\ref{fig: tri spins}. 

Naturally, misaligned triangular spins have another  
relaxation channel, whereby a monopole moves from one kagome plane to the 
next, by flipping the triangular spin without incurring an energy 
barrier. This process is controlled by the time scale for a monopole 
to come by, set by (among other items) the monopole hopping barrier within a 
kagome plane (see main text). 
For $H = 0.4$~Tesla, said barrier is $4.48 H \simeq 1.8$~K, which is clearly 
less efficient than the former (and indeed in worse agreement with the data). 

It is perhaps more interesting to investigate the triangular spin relaxation 
at lower fields, $H \lesssim 0.32$~Tesla, yet large enough so that the 
equilibrium density of reversed triangular spins is negligibly small at the 
temperatures of interest ($H \gtrsim 0.2$~Tesla). 

In this case, flipping a triangular spin and separating the two resulting 
monopoles leads to an overall increase in the energy of the system 
\textit{unless} the monopoles (which live on separate kagome planes) 
eventually meet other opposite monopoles and annihilate. 
This process likely incurs a large barrier due to the Coulomb energy that 
needs to be overcome in order to separate oppositely charged monopoles to 
large distances. 

It would be therefore natural to expect that triangular spin relaxation 
for $0.32 \gtrsim H \gtrsim 0.2$~Tesla proceeds via stray monopole motion, 
which faces a rather low barrier to hopping 
within kagome planes ($4.48 H \lesssim 1.3$~K for $H \lesssim 0.3$~Tesla). 
However, contrary to the field range $0.42 \gtrsim H \gtrsim 0.32$~Tesla, 
where the triangular spins relax whilst plenty of free monopoles are available 
in the system, we clearly see in 
Fig.~3 
that the triangular spin relaxation for $0.32 \gtrsim H \gtrsim 0.2$~Tesla 
takes place in a regime where essentially all remaining monopoles are frozen 
into noncontractible pairs! 
Indeed, in order to collapse the long time decay of the triangular spin density 
in the left and middle panel of Fig.~\ref{fig: tri spins} we ought to invoke 
energy barriers of the order of $3-4$~K, which are entirely inconsistent with 
free monopole hopping barriers $\lesssim 1.3$~K. 

In this regime, it appears that separating monopole pairs to large distances 
is unavoidable, whether the pair originates from the reversal of a triangular 
spin or from an existing noncontractible pair frozen in the system. 
Estimating the corresponding energy barrier in this case is a tall order, 
as it depends on the required separation distance (which is in turn related to 
the density of reversed spins and noncontractible pairs). 
Moreover, in order to compute the Coulomb energy cost to separate two charges 
to said distance, one needs to take into account possible screening effects 
that are expected in a Coulomb liquid~\cite{Castelnovo2011}. 

Although this is beyond the scope of the present manuscript, it is 
intriguing to speculate that this scenario might in fact 
allow to directly probe the long range nature of the Coulomb interations by 
measuring the energy scale needed to collapse the long-time 
triangular spin density decay. Although less straightforward to measure than 
the overall magnetisation of the sample, it may be possible to access 
experimentally the magnetisation of the triangular planes alone using neutron 
scattering measurements in the non-spin-flip channel in the KagIce regime. 
NMR experiments can also in principle pick out the triangular sites, 
as they are symmetry distinct from the kagome ones in a field.

Here we limit ourselves to one further observation. If the monopole pair to 
be separated originates from a triangular spin flip, one expects a negative 
Zeeman energy contribution, whereas if the monopoles come from a 
noncontractible pair, the Zeeman contribution is positive. 
In Fig.~\ref{fig: tri spins} (left and middle panel) we observe 
a larger barrier for $H = 0.2$~Tesla ($4.0$~K) than for $H = 0.3$~Tesla 
($3.2$~K), which suggests 
that the former process is the one responsible for the long time decay of 
the reversed triangular spin density. 
%
%

\section{
Zero-field quenches to spin ice
        }
In this section we discuss the case of quenches down to zero field, 
driving the system from the SatIce directly to the spin ice regime. 
This is experimentally relevant because the characterisation of the behaviour 
of the system following the quench can then be done using measurements in 
zero field, a scenario that is for instance more suitable to using NMR 
techniques. 

\begin{figure}
\hspace{-0.2 cm}\includegraphics[width=0.45\textwidth]{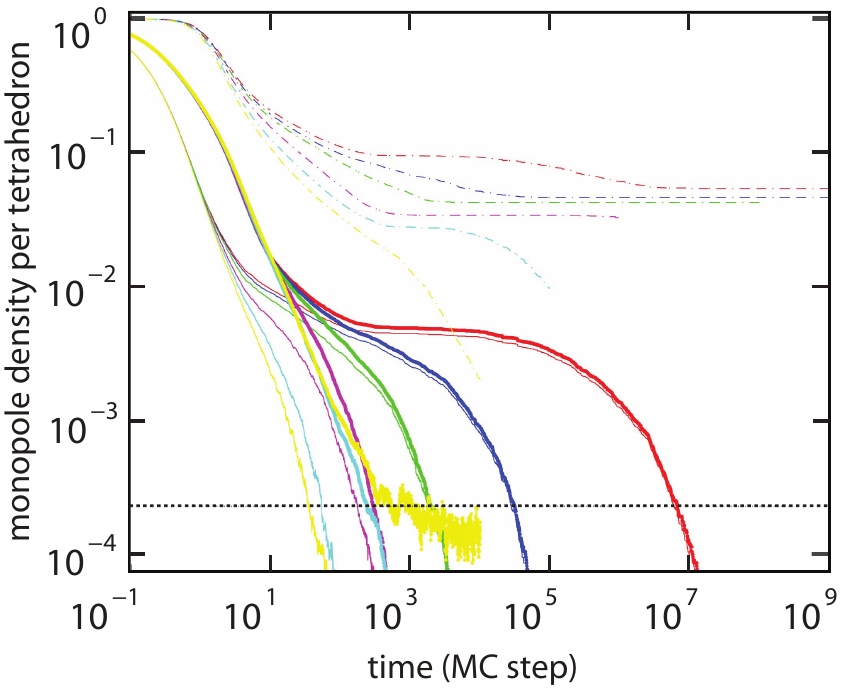}
\caption{\label{fig: Hq=0 decay} 
Monopole density (thick lines), density of triangular spins pointing in the 
direction of the initial magnetization (thin dot-dashed lines), and density of 
noncontractible pairs (thin solid lines) from Monte Carlo simulations 
for a system of size $L=8$, field $H=0$ and temperatures 
$T=0.1,0.2,0.3,0.4,0.5,0.6$~K (red, blue, green, magenta, cyan, and yellow). 
All triangular spins are initially polarised in the direction of the 
the applied field and their initial density is $1$. As time progresses, 
some of the triangular spins reverse, as shown by the decrease in the density 
(magnified by a factor of 10 for visualization purposes). 
The lower the temperature, the larger the density of polarised triangular 
spins that remain `frozen in' in the system. 
}
\end{figure}

The simulation results in Fig.~\ref{fig: Hq=0 decay} show that 
the intermediate-time regime, where the noncontractible pairs are only a 
small fraction of the total monopole density, is essentially absent in 
quenches from SatIce to spin ice. 

The overall behaviour resembles closely the one following a thermal 
quench~\cite{Castelnovo2010}. The two protocols share the feature 
of being quenches from a monopole-rich phase to a spin ice phase with 
low equilibrium monopole density. The only difference is that in one case 
the high-monopole-density phase is a fully-ordered ionic crystal of monopoles 
while in the other it is a disordered Coulomb liquid.~\footnote{Note that the 
ionic crystal state and the disordered plasma state differ not all that much 
in their long range correlations, because long range 
Coulomb interactions suppress charge fluctuations in the plasma phase, 
and topological kinematic constraints in spin ice strictly forbid volume 
charge accumulation~\cite{Castelnovo2010}.}

Once again, the long time behaviour is controlled by noncontractible pairs. 
However, since the quench is to spin ice rather than to kagome ice, the 
triangular spins are not fully polarised in the final state (indeed, there 
is no longer a distinction between kagome and triangular spins 
in zero field). As a result, 
the noncontractible pairs are allowed to decay by direct annihilation of their 
positive and negative monopoles, say, after they hop around a hexagonal 
loop on the lattice. This process has been thoroughly described in 
Ref.~\cite{Castelnovo2010} and has a system-size-\emph{in}dependent 
energy barrier due to the cost of separating two Coulomb interacting 
monopoles up to third-neighbour distance. 

We verified that the Poissonian decay model proposed in 
Ref.~\cite{Castelnovo2010} is in qualitative agreement with the long-time 
behaviour of the monopole density following a field quench to zero field. 
However, a detailed comparison of thermal and zero field quenches would 
require an extensive campaign of low temperature MC simulations which we 
leave for future study.
%
%

\section{
Finite-rate field ramps
        }
Fast changes in the applied magnetic field are experimentally more manageable 
than correspondingly fast changes in temperature. For example, they are not 
limited by the sample heat capacity and thermal conductivity/contact issues. 
Moreover, spin ice samples are insulators and their inductance is negligible. 

However, a sudden change in magnetic field in an experimental setting 
designed to keep the sample at sub-Kelvin temperatures is nonetheless a 
tall order. 
Magnetic field sweeps with a superconducting magnet (as in a typical NMR 
setting) usually achieve no more than $0.8$~Tesla/min. 
Opening the circuit in a solenoid would yield a faster field change, but 
of course this could generate large amounts of 
heat that the refrigerator is then unable to dispense with quickly enough. 

Other techniques that promise to reach larger sweep rates include physically 
moving the sample relative to the magnet (a permanent magnet can be used 
given the relatively small fields involved, $H \lesssim 1$~Tesla); 
or using magnetic field pulses instead of stepping the magnetic field. 

Either way, rates large enough to approximate a 
quench sudden on the time scale of a single spin flip, $\sim 1$~ms, 
will require substantial experimental effort. It is therefore important 
to run field-sweep simulations and compare what part of the out-of-equilibrium 
phenomenology discussed in this paper persists for lower sweep rates.

\begin{figure}
\hspace{-0.0 cm}\includegraphics[width=0.4\textwidth]{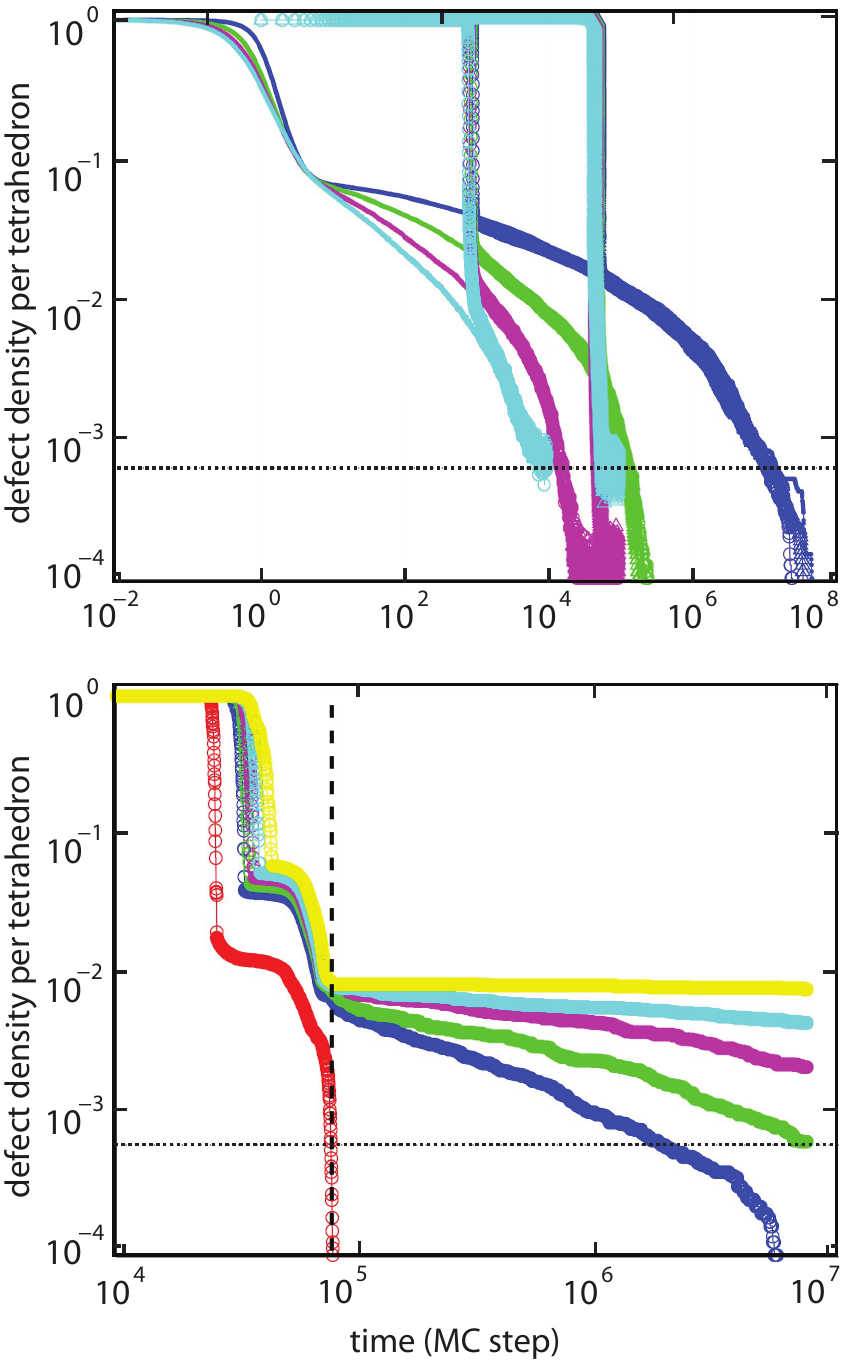}
\caption{\label{fig: field sweeps} 
Top panel: 
Monopole density as a function of time in simulations of a system of size 
$L=6$ with initial and final field values of $H=2$~Tesla and $H=0.5$~Tesla, 
respectively, from three sets of data (from left to right): 
(i) field quench; (ii) field sweep at constant rate over $10^3$~MC steps; 
and (iii) field sweep at constant rate over $6\,10^4$~MC steps. 
Different colours correspond to different temperatures, 
$T=0.2,0.3,0.4,0.5$~K (blue, green, magenta, and cyan). 
Bottom panel: 
Monopole density as a function of time in field sweep simulations from 
$H=1.0$~Tesla down to $H=0$~Tesla at the constant rate of $0.8$~Tesla/min, 
for $T=0.2,0.1,0.09,0.08,0.07,0.05$~K 
(red, blue, green, magenta, cyan, and yellow). 
The vertical dashed line indicates the time where the field reaches its 
final value $H=0$~Tesla. 
}
\end{figure}

In the left panel of Fig.~\ref{fig: field sweeps} we compare the monopole 
density as a function of time from three different protocols: 
(i) an instantaneous quench from SatIce to $H=0.5$~Tesla; 
(ii) a constant rate sweep where the field is lowered from $H = 2$~Tesla at 
time $t=0$ down to $H = 0.5$~Tesla at time $t = 10^3$~MC steps 
(i.e., approximately $1$~s in real time), and it is 
held constant thereafter; 
and 
(iii) a constant rate sweep where the field reaches its final value 
$H = 0.5$~Tesla at time $t = 6\,10^4$~MC steps (i.e., approximately $1$~minute 
in real time). 
Similar results arise for other values of the final applied field. 
It is clear that the long-time regime remains accessible in field sweep 
measurements. As a matter of fact, the field sweep curves merge with their 
quench counterpart promptly after the field has reached its final value. 
Therefore, sufficiently fast field-sweeps can still be used to prepare spin 
ice in a quasi-static monopole-rich phase 
(equivalent to a frozen, over-ionised electrolyte), 
where direct detection of the monopoles might be within reach of zero-field 
NMR measurements. 

In the right panel of Fig.~\ref{fig: field sweeps} we show the behaviour of 
the monopole density in field sweeps down to zero field (at the constant 
rate of $0.8$~Tesla/min, starting from $H = 1$~Tesla). 
Remarkably, even for such slow field sweeps, it is only a matter of 
reaching a sufficiently low temperature ($T \lesssim 1$~K) before the system 
enters a long-lived metastable state where the density of monopoles remains 
much larger than in thermal equilibrium. 



\end{document}